\documentclass{nature_hack}

\newcommand{\araa}{Ann. Rev. Astron. Astrophys.}
\newcommand{\apj}{Astrophys. J.}
\newcommand{\apjs}{Astrophys. J. Suppl.}

\newcommand{\aap}{Astron. \& Astrophys.}
\newcommand{\aj}{Astron. J.}
\newcommand{\mnras}{Mon. Not. R. astr. Soc.}
\newcommand{\pasp}{Publications of the Astronomical Society of the Pacific}
\newcommand{\pasj}{Publications of the Astronomical Society of Japan}
\newcommand{\nat}{Nature}
\newcommand{\aaps}{Astron. Astrophys. Suppl. Ser.}
\newcommand{\zap}{Zeitschrift fuer Astrophysik}
\usepackage{bm}
\usepackage{hyperref}

\newcommand{\apss}{Astrophysics \& Space Science}
\newcommand{\prd}{PRD}
\newcommand{\be}{\begin{equation}}
\newcommand{\ee}{\end{equation}}
\usepackage{deluxetable}

\renewcommand{\baselinestretch}{1.5}

\newcommand{\flux}{\,ergs\,cm$^{-2}$\,s$^{-1}$}
\newcommand{\pflux}{\,ph\,cm$^{-2}$\,s$^{-1}$}

\newcommand{\cm}{\,cm$^{-2}$}
\newcommand{\nh}{$N_\mathrm{H}$}

\newcommand{\bv}{B-V}
\newcommand{\nova}{ASASSN-16ma}
\newcommand\fs{\mbox{$.\!\!^{\mathrm s}$}}%
\newcommand\arcdeg{\mbox{$^\circ$}}%
\newcommand\arcmin{\mbox{$^\prime$}}%
\newcommand\farcs{\mbox{$.\!\!^{\prime\prime}$}}%
\newcommand\farcm{\mbox{$.\!\!^{\prime}$}}%
\usepackage{amsmath}
\usepackage{amssymb}
\usepackage{color}
\usepackage{caption}
\usepackage{placeins}
\usepackage{graphicx}
\usepackage{alphalph}

\title{A Nova Outburst Powered by Shocks}

\author{Kwan-Lok Li$^{1*}$, Brian D. Metzger$^{2*}$, Laura Chomiuk$^{1*}$, Indrek Vurm$^3$, Jay Strader$^1$, Thomas Finzell$^1$, Andrei M.~Beloborodov$^{2}$, Thomas Nelson$^{4}$, Benjamin J. Shappee$^{5}$, Christopher S. Kochanek$^{6,7}$, Jos\'e L. Prieto$^{8,9}$, Stella Kafka$^{10}$, Thomas W.-S. Holoien$^{6,7}$, Todd A. Thompson$^{6,7}$, Paul J. Luckas$^{11}$, \& Hiroshi Itoh$^{12}$}

\begin{document}

\maketitle

\begin{affiliations}
 \item Department of Physics and Astronomy, Michigan State University, East Lansing, MI, 48824, USA
 \item Department of Physics and Columbia Astrophysics Laboratory, Columbia University, New York, NY, 10027, USA
 \item Tartu Observatory, T\~oravere, 61602 Tartumaa, Estonia
 \item Department of Physics and Astronomy, University of Pittsburgh, Pittsburgh, PA, 15260 USA
 \item Hubble, Carnegie-Princeton Fellow, Carnegie Observatories, 813 Santa Barbara Street, Pasadena, CA, 91101 USA
 \item Department of Astronomy, The Ohio State University, 140 West 18th Avenue, Columbus, OH, 43210 USA
 \item Center for Cosmology and Astro-Particle Physics, The Ohio State University, 191 West Woodruff Avenue, Columbus, OH, 43210 USA
 \item N\'ucleo de Astronom\'ia de la Facultad de Ingenier\'ia y Ciencias, Universidad Diego Portales, Av. Ej\'ercito 441, Santiago, Chile
 \item Millennium Institute of Astrophysics, Santiago, Chile
 \item American Association of Variable Star Observers, Cambridge, MA, 02138 USA
 \item International Centre for Radio Astronomy Research, The University of Western Australia, 35 Stirling Highway, Crawley, Perth, WA, 6009 Australia
 \item Variable Star Observer’s League in Japan, 1001-105 Nishiterakata, Hachioji, Tokyo, 192-0153 Japan
\end{affiliations}

\newpage
\begin{abstract}
Classical novae are runaway thermonuclear burning events on the surfaces of accreting white dwarfs in close binary star systems, sometimes appearing as new naked-eye sources in the night sky\cite{2003cvs..book.....W}. The standard model of novae predicts that their optical luminosity derives from energy released near the hot white dwarf which is reprocessed through the ejected material\cite{1978ARA&A..16..171G,1983PASJ...35..507K,1998PASP..110....3G,2012clno.book.....B}. Recent studies with the Fermi Large Area Telescope have shown that many classical novae are accompanied by gigaelectronvolt $\bm{\gamma}$-ray emission\cite{2014Sci...345..554A, 2016ApJ...826..142C}. This emission likely originates from strong shocks, providing new insights into the properties of nova outflows and allowing them to be used as laboratories to study the unknown efficiency of particle acceleration in shocks. Here we report $\bm{\gamma}$-ray and optical observations of the Milky Way nova ASASSN-16ma, which is among the brightest novae ever detected in $\bm{\gamma}$-rays. The $\bm{\gamma}$-ray and optical light curves show a remarkable correlation, implying that the majority of the optical light comes from reprocessed emission from shocks rather than the white dwarf\cite{2014MNRAS.442..713M}. The ratio of $\bm\gamma$-ray to optical flux in ASASSN-16ma directly constrains the acceleration efficiency of non-thermal particles to be $\bm{\sim0.005}$, favouring hadronic models for the $\bm{\gamma}$-ray emission\cite{2015MNRAS.450.2739M}. The need to accelerate particles up to energies exceeding 100 gigaelectronvolts provides compelling evidence for magnetic field amplification in the shocks.
\end{abstract}

\nova\ (a.k.a. PNV J18205200$-$2822100, Nova Sgr 2016d, and V5856 Sgr) is an optical transient source in the constellation Sagittarius, discovered by the All Sky Automated Survey for SuperNovae (ASAS-SN\cite{2014ApJ...788...48S}), on 25.02 October 2016 UT\cite{2016ATel.9669....1S} (a corresponding Modified Julian Day of MJD 57686.02) and identified as a normal classical nova with optical spectroscopy\cite{2016ATel.9678....1L,2016ATel.9849....1R}. 
The optical light curve of the nova after its discovery shows three distinct phases (Figure~1). In Phase I, the nova slowly rose to $m_V\sim8$ mag over two weeks. It then showed a rapid brightening by a factor of $\sim 10$ over just two days (Phase II), reaching a naked-eye peak visual magnitude of 5.4 (MJD 57700). This was followed by a relatively stable decline lasting for several weeks (Phase III; see Figure~1 and Methods). 

Immediately following the optical peak, our Fermi target-of-opportunity (ToO) observation detected strong $\gamma$-ray emission from the nova with a very high photon flux of $F_{ph,\gamma}\approx10^{-6}$\pflux\ (Methods). The $\gamma$-ray emission faded rapidly over the next nine days, with only marginal $\gamma$-ray detections in the following week.
This is among the fastest-evolving $\gamma$-ray light curves seen to date from a nova. The optical and $\gamma$-ray light curves are tightly correlated, declining at the same rate and showing a simultaneous dip in the emission around MJD 57705 (Figure~1). The ratio of the $\gamma$-ray to optical luminosity ($\sim 0.002$) remains constant while the $\gamma$-rays are detectable (Figure~1; see also Supplementary Information, SI hereafter). 

The clear correlation between the $\gamma$-ray and optical light in~\nova~leads us to reconsider the standard model for nova optical emission\cite{1978ARA&A..16..171G,1983PASJ...35..507K,1998PASP..110....3G,2012clno.book.....B}. Traditionally, the optical emission from novae is attributed to the outwards diffusion of energy released by nuclear burning on the central white dwarf, resulting in emission close to the Eddington luminosity (the critical luminosity at which the outwards radiation force balances inward gravity). 
The initial rise of the optical light curve is the result of a photosphere expanding at an approximately constant temperature, as much of the released energy goes into expanding the shell. At later times, the observed bolometric luminosity remains approximately constant, with a balance between the receding photosphere and an increasing temperature as the ejecta become optically thin. 

However, this standard picture provides no obvious explanation for why the optical emission should track the evolution of the $\gamma$-ray emitting shocks, as observed in~\nova. Furthermore, we estimate that \nova\ reached a maximum bolometric luminosity of $L_{\rm tot,opt} \approx 10^{39} (d/4.2 \rm{kpc})^2 \,{\rm ergs\,s^{-1}}$ (Methods), which exceeds the Eddington luminosity by roughly an order of magnitude. Super-Eddington luminosities have also been observed for other novae with well constrained distances (e.g., Nova LMC~1988~\#1\cite{1998MNRAS.300..931S}), and are a long-standing mystery in the field of nova research. 

We instead propose that a large fraction of the optical emission in \nova~originates from the same strong shocks responsible for the $\gamma$-ray emission\cite{2015MNRAS.450.2739M}, providing a natural explanation for the synchronized $\gamma$-ray and optical behavior. This picture capitalizes on the widespread observation that novae undergo abrupt transitions from slower to faster moving outflows\cite{1966MNRAS.132..317F,Kato&Hachisu11m}. In \nova, the structure of the H$\alpha$ emission line indicates
an acceleration of the nova outflow from $\lesssim1100$ km s$^{-1}$ in Phase I (day 2) to 2200 km s$^{-1}$ in Phase III (day 18--25) (Supplementary Figure~2). 
Shortly after the transition, the faster ejecta will rapidly expand and collide with the previous slow ejecta. This drives a shock outwards, 
accelerating particles to relativistic speeds and powering the $\gamma$-ray emission\cite{2014MNRAS.442..713M}. 

Although we observe the shocks directly by their $\gamma$-ray output, most of the total shock power ($L_{\rm sh}\approx 10^{39}\dot{M}_{\rm f4}\,v_{\rm f2}^{2}\,{\rm ergs\,s^{-1}}$, where $\dot{M}_{\rm f}=10^{-4}\dot{M}_{\rm f4}\,M_{\odot}\,{\rm week^{-1}}$ and $v_{\rm f}=2000\,v_{\rm f2}\,{\rm km\,s^{-1}}$ are the mass-loss rate and the velocity of the fast ejecta, respectively; SI) is radiated as thermal X-rays of temperature $kT\sim10$~keV. However, these X-rays are strongly attenuated by the dense slow ejecta ahead of the shocks and will ultimately escape as UV/optical light. This reprocessed emission can dominate the observed optical luminosity of $L_{\rm opt} \sim 10^{38-39}\,{\rm ergs\,s^{-1}}$ in Phases II and III\cite{2015MNRAS.450.2739M}, and is consistent with the X-ray upper limit derived from Swift telescope data around optical maximum (Methods and SI). It is also consistent with the observed constant ratio of the $\gamma$-ray and optical luminosities (Figure~1), assuming the efficiency of relativistic particle acceleration also remains constant in time. 
In our model, the duration of the optical rise in Phase II is the time required for this reprocessed radiation to diffuse through the dense, optically-thick slow outflow. The $\gamma$-rays are temporarily attenuated by inelastic electron scattering in the same outflow before their detection at the end of Phase II (SI).

Key details of the microphysics of the non-thermal emission can be inferred from the $\gamma$-ray spectral energy distribution (SED) of \nova. 
Based on a detailed model for the $\gamma$-ray emission from radiative shocks\cite{2016arXiv161104532V} (see SI for details), we fit the SED in the 100 MeV to 300 GeV energy range to two competing scenarios, the hadronic and leptonic models\cite{2013A&A...551A..37M,2014Sci...345..554A,2015MNRAS.450.2739M}. In the hadronic model, accelerated ions strike other ions, producing pions that decay into $\gamma$-rays. In the leptonic model, the accelerated electrons emit $\gamma$-rays via bremsstrahlung and inverse Compton processes.

Both models give qualitatively reasonable fits (see Figure~2 and SI; for references only, $\chi_\nu^2=3.04/6$ and $3.86/6$ for leptonic and hadronic models, respectively). However, consideration of the derived model parameters lends support to the hadronic model. A high shock magnetization (i.e., $\epsilon_{B} \gtrsim 10^{-4}$, where the magnetic field $B = v_{\rm sh} \sqrt{6\pi\epsilon_{B}\rho}$ and $\rho$ is the upstream density) is required to accelerate particles to energies $\gtrsim 10-100$ GeV sufficient to explain the highest energy $\gamma$-rays, providing strong evidence for magnetic field amplification at the shock (SI). This high magnetization is, however, incompatible with leptonic models, which require $\epsilon_{\rm B} \lesssim 10^{-6}$ to avoid strong synchrotron cooling losses behind the shock (SI). Hadronic models are insensitive to the magnetic field because proton synchrotron cooling is negligible. The hadronic model also naturally predicts a low-energy spectral turnover near the pion rest energy of $\sim 140$ MeV, as observed in \nova. However, reproducing this same spectral shape in leptonic models requires a lower optical seed radiation field for inverse Compton emission than would be present if the shocks were embedded behind a large column of gas, as our light curve models require for independent reasons. Finally, the inferred proton acceleration efficiency in the hadronic model of $\epsilon_p \approx 5\times10^{-3}$ is compatible with the theoretical upper limit from hybrid kinetic shock simulations\cite{2014ApJ...783...91C} ($\epsilon_p \leqslant 0.2$; see SI for further explanation). By contrast, in leptonic models, a large non-thermal electron acceleration efficiency of $\epsilon_{\rm e} \approx 2.5 \times 10^{-3}$ is required, in tension with the value $\epsilon_{\rm e}\sim10^{-4}$ derived by modeling supernova remnant emission\cite{2012A&A...538A..81M} and simulations of particle acceleration in shocks\cite{2015PhRvL.114h5003P}. Taken as a whole, the hadronic model is favoured over the leptonic one. 

We examined the available Fermi-LAT $\gamma$-ray and optical data of all other five Fermi-detected classical novae\cite{2014Sci...345..554A,2016ApJ...826..142C,2016ATel.9699....1L} to explore what evidence exists for the correlation between optical and $\gamma$-ray luminosity clearly observed in \nova\ (V959 Mon 2012 and V407 Lup 2016 are excluded in the analysis because the former has no useful optical data taken when the $\gamma$-ray counterpart was bright\cite{2014Sci...345..554A}, and the later was just marginally detected with a significance of 4.4$\sigma$\cite{2016ATel.9594....1C}). 
Although marginally, the two $\gamma$-ray classical novae, V339 Del and V5855 Sgr, exhibited correlated optical/$\gamma$-ray light curves with significances of $\sim2\sigma$, suggesting that \nova\ may not be an orphan (Supplementary Figure~6). If the particle acceleration efficiency in \nova\ is representative of that in all the $\gamma$-ray novae, then given the $\gamma$-ray/optical luminosity ratios in the other novae\cite{2014Sci...345..554A,2015MNRAS.450.2739M,2016ApJ...826..142C}, we can conclude that a significant fraction of the optical emission in these other events must also be powered by shocks. In V1369 Cen and V5668 Sgr, weak evidence may exist to support the claim insofar as the $\gamma$-ray emission in both events was detected only during epochs of bright optical emission (i.e., $m_V\lesssim5$ mag for V1369 Cen and $m_V\lesssim6$ mag for V5668 Sgr\cite{2016ApJ...826..142C}). In addition, the $\gamma$-ray emission in most other novae begins near the first optical peak\cite{2014Sci...345..554A, 2016ApJ...826..142C}, as expected from shocks developing between the fast and slow flows. 
Prompt $\gamma$-ray observations of future Galactic novae will allow the the universality of this model to be tested. 

Through $\gamma$-ray and optical observations of \nova, we have shown that nova optical emission can be indirectly powered by shocks below the photosphere, thus verifying a prediction first made by Refs.\cite{2014MNRAS.442..713M,2015MNRAS.450.2739M}. 
This discovery challenges the standard model that most UV/optical emission in novae is the result of outwards diffusion of the radiation from the white dwarf\cite{2005ApJ...623..398Y}. In addition, it provides a solution for the long-standing mystery of why many novae exceed the Eddington luminosity\cite{Duerbeck81}---shock driven emission, unlike the hydrostatic atmosphere of the white dwarf, obeys no such luminosity limit (SI). 
Our results also confirm how the dense nova ejecta serve as an effective ``calorimeter'' for relativistic particles\cite{2015MNRAS.450.2739M}: the power and acceleration efficiency of the shock are measured directly from the UV/optical and $\gamma$-ray luminosities, respectively (SI). Given the conditions at the shock required to explain the observed luminosities, coupled with the need to accelerate particles to energies exceeding 100 GeV, we find strong evidence for magnetic field amplification at the shocks (SI), a topic of active debate in other astrophysical settings like supernova remnants\cite{2014ApJ...790...85R}. 

\begin{addendum}
 \item We acknowledge with thanks the variable star observations from the AAVSO International Database contributed by observers worldwide and used in this research. 
We thank the Fermi Science Support Center, supported by the Flight Operations Team, for scheduling the Fermi ToO observations. We also acknowledge the use of public data from the Fermi and the Swift data archives. 
We thank Las Cumbres Observatory and its staff for their continued support of ASAS-SN. ASAS-SN is funded in part by the Gordon and Betty Moore Foundation through grant GBMF5490 to the Ohio State University. ASAS-SN is supported by NSF grant AST-1515927. Development of ASAS-SN has been supported by NSF grant AST-0908816, the Center for Cosmology and Astro Particle Physics at the Ohio State University, the Mt. Cuba Astronomical Foundation, and by George Skestos. 
We also acknowledge useful discussions with J. Linford, A. Mioduszewski, K. Mukai, J. Sokoloski, K. Stanek, and J. Weston, that have greatly improved the quality of the paper. 
This work was partially supported by Fermi GI grant NNX14AQ36G. J.S.~acknowledges support from the Packard Foundation.
L.C.~acknowledges a Cottrell Scholar Award from the Research Corporation for Science Advancement. ​
B.J.S. is supported by NASA through Hubble Fellowship grant HST-HF-51348.001 awarded by the Space Telescope Science Institute, which is operated by the Association of Universities for Research in Astronomy, Inc., for NASA, under contract NAS 5-26555. T.W.S.H. is supported by the DOE Computational Science Graduate Fellowship, grant number DE-FG02-97ER25308. 
C.S.K. and T.W.S.H. are supported by NSF grants AST-1515876 and AST-1515927. 

 \item[Author Contributions] K.L.L., B.D.M., L.C., I.V., and J.S. wrote the text. B.J.S, C.S.K., J.L.P., T.W.S.H., and T.A.T., on behalf of the ASAS-SN team, discovered \nova\ and provided the ASAS-SN data. S.K., the director of AAVSO, provided the AAVSO light curves, through the AAVSO International Database. 
P.J.L. observed and provided the ARAS spectroscopic data. 
H.I., on behalf of the VSNET team, provided useful photometric data. 
L.C., K.L.L, and J.S. requested and obtained the Fermi-LAT observations. T.N. requested and obtained the Swift observation. K.L.L. analysed the Fermi LAT, the Swift XRT, the AAVSO, and the VSNET observations. 
J.S. analysed the ARAS observations. 
B.D.M., I.V., and A.M.B. worked on the theoretical interpretation of the data. T.F. contributed to the distance and extinction estimations and provided useful data of V1324 Sco and V339 Del for meaningful comparisons. All authors discussed the results and commented on the final manuscript. 
 \item[Competing Interests] The authors declare that they have no competing financial interests.
 \item[Correspondence] Correspondence and requests for materials should be addressed to K.L.L.~(\href{mailto:liliray@pa.msu.edu}{liliray@pa.msu.edu}), B.D.M.~(\href{mailto:bdm2129@columbia.edu}{bdm2129@columbia.edu }), and L.C.~(\href{mailto:chomiuk@pa.msu.edu}{chomiuk@pa.msu.edu}).
\end{addendum}

\newpage
\begin{figure}
\includegraphics[width=\textwidth]{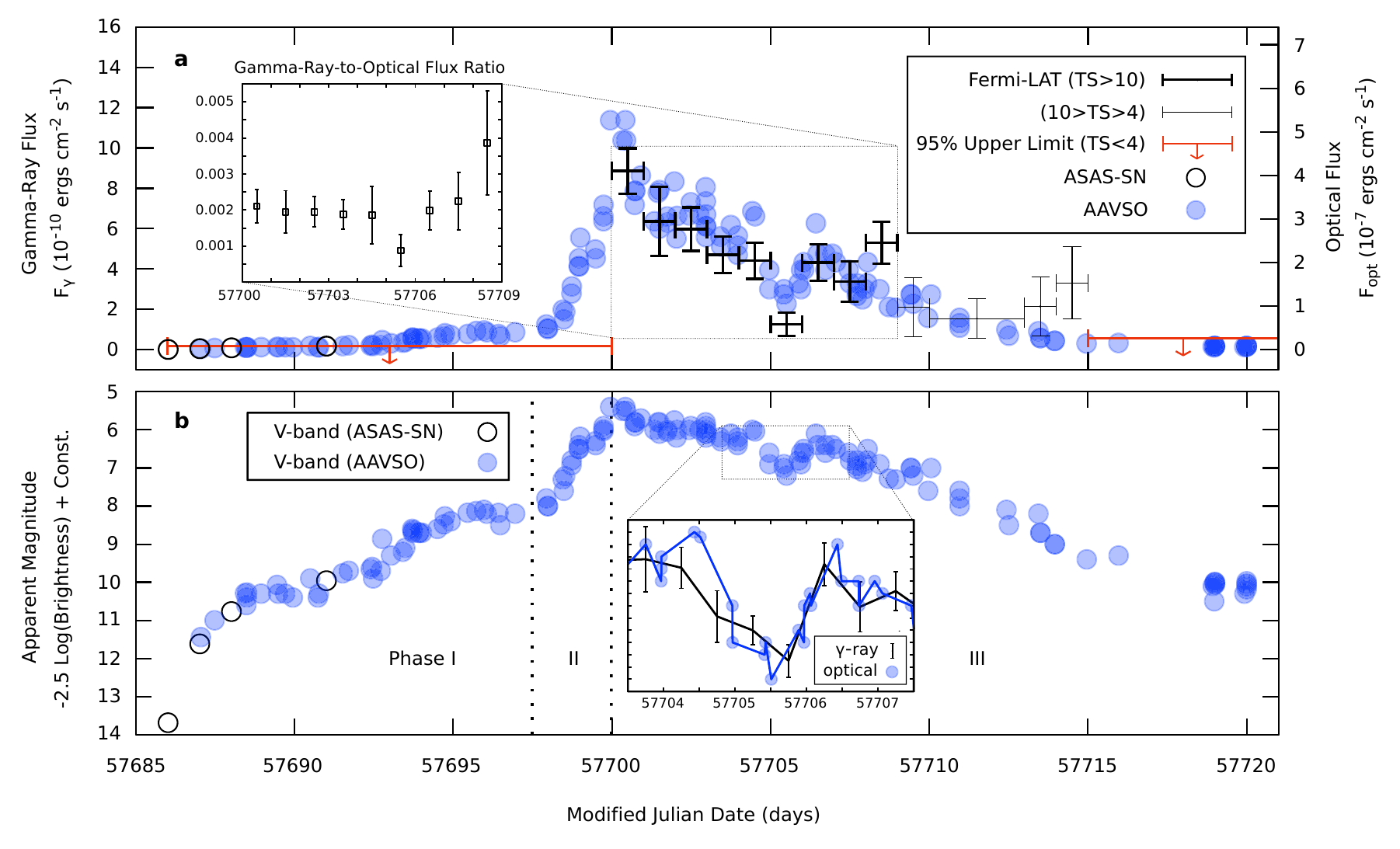}
\centering
\caption{\textbf{Optical and ${\bm\gamma}$-ray light curves track each other}. The top panel (a) shows the $\gamma$-ray (black and gray crosses, and red arrows) and the bolometric (blue and black circles) light curves of \nova\ in flux units of \flux, using observations from Fermi-LAT and ASAS-SN/AAVSO. For the Fermi-LAT light curve, most of the data points are daily binned, while some are combined from several daily bins that have low detection significances (i.e., $<2\sigma$). 
Black and gray colours (with 1$\sigma$ error bars) represent detection significances larger than $\sim3\sigma$ (i.e., $\mathrm{TS}>10$) and between $\sim2$--$3\sigma$ (i.e., $10>\mathrm{TS}>4$), respectively, while 95\% upper limits are indicated by red arrows for bins with detection significance below $\sim2\sigma$ (i.e., $\mathrm{TS}<4$). 
The inset box in (a) shows the $\gamma$-ray to optical flux (luminosity) ratio of \nova, which remained $F_\gamma/F_\mathrm{opt}\sim0.002$ over the whole $\gamma$-ray active period. 
The bottom panel (b) shows the $V$-band light curves of the same optical datasets used in (a), however, on a magnitude (logarithmic) scale, that clearly shows the three different phases of the optical light curve. The inset box in (b) zooms in on the emission dip at MJD 57705 in $\gamma$-rays and optical, which directly show the co-variance of the $\gamma$-ray and optical emission on time-scales as short as 0.5 days. 
 }
\end{figure}

\begin{figure}
\includegraphics[width=\textwidth]{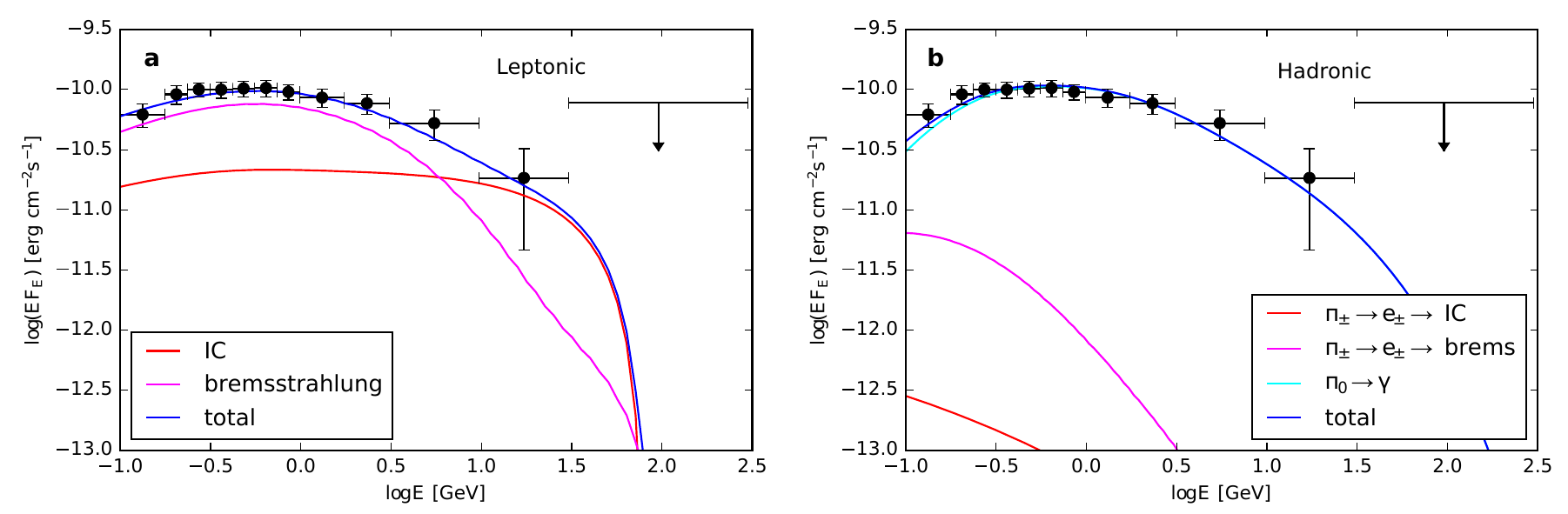}
\centering
\caption{\textbf{The hadronic model is favoured over the leptonic model based on the parameters of the model fits to the Fermi $\gamma$-ray spectrum. }
Leptonic model (a): Electrons are injected at the shock with the spectrum
$dN_{\rm e}/d\ln\gamma\propto \gamma^{-q}$, where $q=2$ (equal energy per logarithmic $\gamma$ interval).
Hadronic model (b): 
Proton injection spectrum
$dN_{\rm p}/d\ln(\gamma\beta)\propto (\gamma\beta)^{-q}$, where $q=2.7$.
The shock speed and the total shock luminosity are assumed to be $2000$~km~s$^{-1}$ and the average optical luminosity, respectively. 
The radiative processes forming each spectrum are labeled following the inset for each panel. 
While both models can describe the data within the statistical errors, the hadronic model is favoured, because the leptonic model requires an implausibly high electron acceleration efficiency and cannot produce the $>$10 GeV $\gamma$-ray emission in a self-consistent way (see the main text for details). The reported errors are 1$\sigma$ uncertainties and the upper limits are at 95\% confidence level. 
}
\end{figure}

\FloatBarrier

\newpage
\begin{methods}

\subsection{ASAS-SN Discovery of ASASSN-16ma.}
\nova\ was detected at $m_V=13.69\pm0.02$~mag by the ASAS-SN on 25.02 October 2016 UT with the quadruple 14-cm ``Cassius'' telescope in CTIO, Chile\cite{2016ATel.9669....1S}. A rising light curve (Figure~1) is revealed by the subsequent observations with magnitudes: $m_V=11.62\pm0.01$~mag (day 1), $m_V=10.77\pm0.01$~mag (day 2), and $m_V=9.96\pm0.01$~mag (day 5). 
No object can be detected at the nova position in the previous ASAS-SN observations starting from March 2016 and the last pre-discovery image taken on 20.04 October 2016 places a limiting magnitude of $m_V>17.3$~mag on the progenitor. 

\subsection{American Association of Variable Star Observers (AAVSO): Photometry, Temperature, and Distance.}
AAVSO is an astronomical association of observers, amateur or professional, that provides optical photometric observations of variable sources. From the AAVSO International Database, we downloaded the photometric datasets, including both visual estimates of magnitudes and those quantified by CCD photometry. Most of the AAVSO data are visual magnitudes, which for experienced observers are consistent with standard $V$ magnitudes. Some epochs have CCD photometry, including multi-band measurements in $BV$ that we use to estimate the nova temperature below. The photometry presented here spans the date range MJD 57687--57719. Unless mentioned otherwise, a Galactic reddening of $E(B-V)=0.34$ mag ($A_V=1.06$ mag with $R_V=3.1$) is assumed\cite{2011ApJ...737..103S}.

From the $V$-band light curve, the optical evolution can be divided into three major phases, namely Phase I, II , and III (Figure~1). In Phase I (from about MJD 57687 to 57697), the optical emission rose at a mean rate of $\dot{m}_V\approx-0.3$~mag day$^{-1}$, but with two apparent plateaus in the light curve during which the optical brightness did not change for 2--3 days. In Phase II (from about MJD 57698 to 57700), the nova brightened rapidly by $\dot{m}_V\approx-1.3$~mag day$^{-1}$ and the emission reached a maximum of $m_V=5.4$~mag (or $m_{V,0}=4.3$~mag corrected for extinction) on MJD 57700. In Phase III, the emission started to decrease from the peak at a rate of $\dot{m}_V\approx0.2$~mag day$^{-1}$ until the end of the data set. 
The only clear short-term variability is a $\Delta m_V\approx1$~mag dip around MJD 57705.5 that lasted for 1--2 days. This dip feature is also present with a similar profile in the Fermi-LAT MeV/GeV light curve (Figure~1). 
We noted that the ANS Collaboration also monitored the nova with a 40~cm robotic telescope and all the aforementioned optical features also appear in the ANS light curve\cite{2017arXiv170309017M}. 

Color temperatures ($T_c$) of the nova photosphere over time were estimated with the night sets of $B$- and $V$-band data. While the earliest dataset (MJD 57687, about 1 day after the ASAS-SN discovery) shows a relatively blue color of $(\bv)_0=-0.02$~mag (extinction corrected; equivalent to a color temperature of $T_c=10300$~K\cite{2009aste.book.....K}), the subsequent colors level off in the range of $(\bv)_0=0.16$--0.27~mag ($T_c=7200$--8100~K) with an average $(\bv)_0=0.21$~mag ($T_c=7700$~K) and the latest color on MJD 57704.5 goes redder to $(\bv)_0=0.57$~mag ($T_c=5500$~K; see Supplementary Table~1). 
The observed color near the optical peak, $(\bv)^{\rm max}_0=0.24$, is close to that typically observed for novae\cite{1987A&AS...70..125V}, showing the adopted extinction is reasonable (Supplementary Table~1). 

The decay rate in Phase III was used to estimate the nova distance through the maximum magnitude-rate of decline (MMRD) relation. We use the linear empirical equation of $V$-band absolute magnitude at maximum ($M_V$) and the time (in days) needed to decline two magnitudes from the peak ($t_2$), of
\begin{equation}
M_V=(-11.32\pm0.44)+(2.55\pm0.32)\log t_2, 
\end{equation}
from Downes et al.\cite{2000AJ....120.2007D} to estimate an absolute magnitude of $M_V=-8.8\pm0.5$~mag for $t_2=10$ days, implying a distance of $d=4.2^{+1.2}_{-0.9}$~kpc, which is roughly consistent with $d\approx6.4$~kpc estimated by Munari et al.\cite{2017arXiv170309017M}. We also checked the result with the non-linear version\cite{2000AJ....120.2007D}, 
\begin{equation}
M_V=-8.02-1.23\arctan \Big[\frac{1.32-\log t_2}{0.23}\Big], 
\end{equation}
which gives a consistent result of $M_V=-9.2$~mag and $d=5.1$~kpc. 
Despite the large uncertainties and the uncertain reliability of the MMRD method\cite{2011ApJ...735...94K}, many of our main results, like the acceleration efficiency inferred from the $\gamma$-ray to optical flux ratio, are not sensitive to the distance. We adopted the linear MMRD distance of $d=4.2$~kpc as a reference distance in our analysis but caution that the uncertainties in the distance do not include systematic uncertainties in the MMRD. For example, Kasliwal et al.\cite{2011ApJ...735...94K} found that some novae with $t_2 \sim 10$ days (the observed value for ASASSN-16ma) were up to 2 mag fainter than predicted by the MMRD, which would imply a distance of only 1.7 kpc. At this short distance the inferred luminosity could be consistent with the Eddington luminosity, depending on the mass of the white dwarf. 
Besides the MMRD technique, we assumed the peak brightness of \nova\ equal to the Eddington luminosity for a 1 $M_\odot$ white dwarf and $M_V\approx-8.7$, the mean absolute magnitude of the Galactic novae at peak\cite{2012JAVSO..40..582M}, which infer distances of $d=1.4$~kpc and 4.1~kpc, respectively. The later value is very close to our reference distance of 4.2~kpc. 

\subsection{Variable Star Network (VSNET): Photometry and Temperature. }
VSNET is a collaboration whose members share their observational results (mainly in optical) of variable stars and new transients, including \nova. Supplementary Table~1 presents the VSNET photometric measurements and the inferred color temperatures (using the same approach as for the AAVSO data), which are mostly consistent with the AAVSO result. 

\subsection{Astronomical Ring for Access to Spectroscopy (ARAS): Temperature and H$\bm\alpha$ Line. }
Despite its brightness, only limited optical spectroscopy of ASASSN-16ma was possible owing to its discovery close to the Sun. Spectroscopy was undertaken by Luckas over 8 epochs, from 27 Oct to 18 Nov 2016 (UT) and submitted to the ARAS database for distribution. The observations used an Alpy 600 spectrograph with a low-resolution grism, mounted on a 36 cm telescope. The exposure times varied from around 1 hour (right after discovery) to 10 minutes (near optical peak). The observations were reduced in the standard manner, with wavelength calibration using a Ne arc lamp. The resulting spectra have a resolution of about 11.4 \AA\ FWHM (520 km s$^{-1}$) in the region of H$\alpha$. 

After extinction corrections, we fit a blackbody model to the five spectra with clear photospheric continuum emission (at late times, the spectra are consistent with only line emission).
The inferred temperatures are consistent with those extracted from the AAVSO and VSNET photometry (Supplementary Table~1). In addition, we detect the H$\alpha$ emission line in all 8 epochs of spectroscopy, with the profile changing from an unresolved narrow line with a pair of $\pm$1100 km/s wings, to a P Cygni profile, and finally a broad line with wings extending to $\pm$2200 km/s (Supplementary Figure~2). 

\subsection{Pre-Nova Observations from the Optical Gravitational Lensing Experiment (OGLE) and the VISTA Variables in the Via Lactea (VVV) survey. }
While a faint near-IR VVV source detected in 2010 was first proposed to be the progenitor\cite{2016ATel.9680....1S}, it has been shown that the VVV source is $1\farcs6$ from the nova location and cannot be the true progenitor\cite{2016ATel.9683....1M}. In fact, no progenitor source has yet been detected to $m_I>22$~mag, constrained by the \textit{OGLE deep template image}\cite{2016ATel.9683....1M}. Taking the extinction $A_I=0.63$~mag and the MMRD distance $d=4.2$ kpc into account, the absolute magnitude of the progenitor is $M_{I,0}>8.3$~mag, indicating that the binary companion of the nova system is likely a main-sequence star, probably a M-type dwarf. 

\subsection{Bolometric Correction of the Optical Light Curve. }
For a comprehensive comparison to the Fermi $\gamma$-ray data, bolometric corrections\cite{1967ZA.....67..415W} ($BC$) were estimated for 16 epochs (i.e., MJD 57687.0 to 57704.5) based on the temperatures inferred from the AAVSO, VSNET photometry, and the ARAS spectra. 
Individual corrections were estimated using: (i) the first measurement of $BC=-0.38$~mag for epochs before MJD 57687.0; (ii) interpolation of the measurements for epochs between MJD 57687.0 and 57704.5 (i.e., $BC=-0.11$ to $-0.19$~mag); or (iii) the last measurement of $BC=-0.19$~mag for the later epochs. 
Although the adopted $BC$ values may not be ideal for all epochs (i.e., early-Phase I and late-Phase III), the corrections are sufficient for our purposes. 

\subsection{Fermi Large Area Telescope (LAT). }
We used Fermi-LAT PASS 8 observations taken from 2016 October 25 (Mission Elapsed Time (MET): 499046404s; the ASAS-SN discovery date) to November 28 (MET: 502043986s) within a circular region of interest (ROI) of 20 degrees radius around the nova optical position at $\alpha\mathrm{(J2000)}=18^\mathrm{h}20^\mathrm{m}52\fs12$, $\delta\mathrm{(J2000)}=-28\arcdeg 22\arcmin 13\farcs52$\cite{2016ATel.9669....1S}. For nearly the entire time from October 25 to November 16, Fermi-LAT was operated in a target-of-opportunity (ToO) Galactic Center-biased survey mode for the nova field. This mode was serendipitously active from the initial discovery of ASASSN-16ma, as the ToO observations were originally triggered for another nearby classical nova (V5855 Sgr; observation number: 090603-1-1; PI: Laura Chomiuk). Fermi-LAT resumed observations in its regular survey mode after November 16. The period of highest sensitivity observations covers nearly all of Phases I and II and the first 8 days of Phase III, ending on MJD 57708.

\texttt{Fermi Science Tools} (version v10r0p5) were used to analyze the LAT data (100~MeV--300~GeV) by following the on-line data analysis threads in the Fermi Science Support Center (FSSC). 
We first constructed an emission model of the event data by considering all cataloged sources within 30 degrees from the nova (i.e., ROI + an extra 10 degrees) in the LAT 4-year Point Source Catalog (3FGL\cite{2015ApJS..218...23A}) with the diffuse background components of Galactic diffuse emission (\texttt{gll\char`_iem\char`_v06}) and extragalactic isotropic diffuse emission (\texttt{iso\char`_P8R2\char`_SOURCE\char`_V6\char`_v06}). 
As the nearby sources are too faint to significantly affect the result (in fact, \nova\ was the dominant source within 5 degrees of the field during the time of interest), we only allowed the intensities (normalization) of two closest 3FGL sources (i.e., 3FGL J1816.2$-$2726 and 3FGL J1823.7$-$3019; 3 degrees within the nova) to vary and fixed all the other spectral parameters of the 3FGL sources in the model file. 
In addition, we noticed from our trial runs that the best-fit extragalactic isotropic diffuse emission would drop to 60\% of the regular level if it is not fixed. So, we also fixed its normalization to unity (i.e., default value) to avoid overestimating the $\gamma$-ray flux of \nova. 

An initial analysis was performed with the task, \texttt{like\char`_lc} (version 1.72; developed by R. Corbet), which is a python script for generating LAT light curves using the unbinned likelihood analysis method. With a simple power-law model fit to the $\gamma$-ray spectrum ($dN/dE\propto E^{-\Gamma_\gamma}$ with a fixed $\Gamma_\gamma=2.1$, the best-fit value found by the binned likelihood analysis described later in this section), we extracted a 1-day binned light curve, in which significant $\gamma$-ray emission was detected starting from 2016 November 8 (MJD 57700, also the date of the optical peak) with a daily $\mathrm{TS}=200$ (equivalent to a detection significance of $\sqrt{\mathrm{TS}}\approx14\sigma$) and a photon flux of $F_{ph,\gamma}=(1.04\pm0.13)\times10^{-6}$\pflux\ (100~MeV--300~GeV). 
The $\gamma$-ray flux then decreased with $F_{ph,\gamma}\propto (t-t_0)^{-0.33\pm0.04}$ (where $t_0={\rm MJD} 57700$, the $\gamma$-ray onset) and dropped below the Fermi-LAT $3\sigma$ detection limits (i.e., $\mathrm{TS}<10$) on 2016 November 17. After this, only weak $\gamma$-ray signals with $\sim2\sigma$ were occasionally detected. We therefore 
conclude that the detectable $\gamma$-ray emission lasted for 9--15 days total (Figure~1 and Supplementary Table~2). In addition to the overall trend of decreasing $\gamma$-ray flux, we clearly detect a dip around MJD 57705.5. To check the spectral index of the dip, we tried to free the daily photon index, but the dip duration is too short for meaningful constraints. However, a general softening trend over the whole $\gamma$-ray active phase is marginally observed (Supplementary Figure~5). This could support a weakening attenuation of the $\gamma$-ray emission as the system evolves (see SI). 

Besides the likelihood light curve, a 0.5-day binned light curve extracted by aperture photometry using the task \texttt{aperture} (version 1.53; developed by R. Corbet) and a circular region of 1 degree radius was examined. Despite the non-subtracted background, the aperture light curve is mostly consistent with the likelihood one, with the same dip feature and a similar decline. With a finer resolution, the aperture light curve confirms that the $\gamma$-ray onset started with the optical peak and also shows that the dip profiles in optical and $\gamma$-rays are very similar to each other in structure (Figure~1). 

Using all the Fermi-LAT observations with daily $\mathrm{TS}>10$, we performed a stacked spectral analysis using the binned likelihood method. For the same ROI, emission model, and energy range, we modeled the data with a simple power-law resulting in $\mathrm{TS}=635$, $\Gamma_\gamma=2.11\pm0.05$, and $F_{ph,\gamma}=(5.9\pm0.5)\times10^{-7}$\pflux\ (errors are statistical only). We found that the fit could be improved significantly ($\sim3 \sigma$ using the likelihood ratio test) by introducing an exponential cutoff ($dN/dE\propto E^{-s}\exp(-E/E_c)$, where $E_c$ is the cutoff energy), to find $\mathrm{TS}=644$, $s=1.86\pm0.11$, $E_c=5.9\pm2.6$~GeV, and $F_{ph,\gamma}=(5.4\pm0.5)\times10^{-7}$\pflux\ (Supplementary Table~3). 
We also carried out a time-resolved spectral analysis before and after the dip, which shows the spectrum getting softer from $\Gamma_\gamma=2.05\pm0.06$ to $2.22\pm0.10$, confirming the softening trend seen in the unbinned likelihood light curve. Adding an exponential energy cutoff does not significantly improve the after-dip spectral fit, in contrast to the before-dip and the overall fits (Supplementary Table~3). This could be caused by an unknown intrinsic spectral change or, more likely, the lower signal-to-noise ratio of the after-dip data. 
To produce a SED for detailed modeling, we also manually split the data into 12 energy bins and ran binned likelihood analyses (Figure~2; see SI for details of the spectral model). 

We refined the position of the $\gamma$-ray source using the task \texttt{gtfindsrc} with the ROI decreased from 20 to 10 degrees to save computational time. The optimized coordinates are $\alpha\mathrm{(J2000)}=18^\mathrm{h}20^\mathrm{m}57\fs87$, $\delta\mathrm{(J2000)}=-28\arcdeg 21\arcmin 40\farcs9$ (95\% error circle radius of $2\farcm7$). The nova is $1\farcm4$ from this position and hence well inside the Fermi error circle.

\subsection{Gamma-ray and Optical Correlation in \nova. }
To estimate the significance of the correlation of \nova, we computed the Pearson correlation coefficients of the AAVSO $V$-band light curve with (i) the Fermi aperture light curves ($r_{a}$), and (ii) the Fermi-LAT likelihood light curves ($r_{l}$). Since the $\gamma$-ray and optical light curves were sampled differently, the optical data was linearly interpolated to pair up with the $\gamma$-ray data. For the Fermi-LAT data with daily $\mathrm{TS}>10$, the Pearson coefficients are $r_{a}=0.75$ ($dof=16$) and $r_{l}=0.86$ ($dof=7$), which correspond to two-tailed $p$-values of $3.7\times10^{-4}$ ($3.6\sigma$) and $3.1\times10^{-3}$ ($3\sigma$), respectively. 
Note that the above significances are very conservative, as the $\gamma$-ray observations with low detection significances ($\mathrm{TS} < 10$) and the temporal coincidence of the $\gamma$-ray and the optical peaks are not considered. 

\subsection{The Peak/Dip Offsets between the Gamma-Rays and Optical Light Curves. }
Thanks to the continuous monitoring by Fermi, the $\gamma$-ray peak (dip minimum) is well constrained on MJD $57700.05\pm0.05$ (MJD $57705.75\pm0.25$) by a 0.1-day (0.5-day) binned light curve. For the AAVSO light curve, owing to the relatively low sampling rate, the optical peak (dip minimum) falls in a wide range of MJD 57699.95--57700.34\cite{2017arXiv170309017M} (MJD 57705.50--MJD 57705.96). After subtractions, the offsets (positive for a $\gamma$-ray delay) are from $-$8 to +4 hours and from $-$11 to +12 hours, for the peak and dip, respectively. 

\subsection{Swift XRT. }
During the $\gamma$-ray active phase, the visibility of ASASSN-16ma was limited for all satellite X-ray observatories due to the Sun angle constraint. Thus, there is only one short (2~ks) X-ray observation for \nova\ taken by Swift XRT in the Windowed Timing mode (WT) on MJD 57701 (2016 November 9; one day after the $\gamma$-ray onset). 
As the data is very noisy below 1~keV, only 1--10~keV band was used to search for the X-ray counterpart. At a 3$\sigma$ threshold, ASASSN-16ma was not detected with a 95\% upper limit\cite{1991ApJ...374..344K} of $< 4 \times10^{-13}$\flux\ (assuming an absorbed power-law with $\Gamma_X=2$ and foreground absorption\cite{2005A&A...440..775K} of \nh\ $=1.33\times10^{21}$\cm; absorption corrected). This is consistent with a 1--10 keV upper limit of $L_X \lesssim 10^{33}$ ergs s$^{-1}$ at the adopted distance. About 5 months later, we requested another 2.4~ks Swift XRT observation, taken in the Photon Counting mode (PC) on 2017 March 23 (MJD 57835). The nova was still undetected with an upper limit\cite{2007A&A...469..379E} of $< 2 \times10^{-13}$\flux\ (or $4 \times 10^{32}$ ergs s$^{-1}$; 0.3--10 keV). 

\subsection{Data Availability. }
The Fermi-LAT, AAVSO, VSNET, and ARAS data that support the findings of this study are available in/from the LAT Data Server (\url{https://fermi.gsfc.nasa.gov/ssc/}), the AAVSO data archive (\url{https://www.aavso.org/data-download}), the VSNET official page (\url{http://www.kusastro.kyoto-u.ac.jp/vsnet/}), and 
the ARAS spectral database (\url{http://www.astrosurf.com/aras/Aras_DataBase/DataBase.htm}),
respectively.

\end{methods}


\newpage
\setcounter{figure}{0}   
\renewcommand{\baselinestretch}{1}
\title{\Large\textbf{Supplementary Information}}

\maketitle

\section{General Descriptions of the Fast-Slow Outflows Collisions in Novae}

Novae all occur in accreting white dwarf (WD) systems. A classical nova system\cite{Townsley&Bildsten05} typically consists of a WD primary of mass $M_{\rm wd} \approx 1.0 M_{\odot}$ and a main sequence (MS) secondary of mass $M_{\star} = 0.5M_{\odot}$ in a binary of period $P \approx 4-7$ hours, separated by $a \approx 10^{11}$ cm. 
The secondary fills its Roche Lobe, losing matter through the inner Lagrangian point that is eventually accreted onto the WD. The material lost through Roche Lobe Overflow (RLOF) forms a thin, degenerate layer on the WD. Once the temperature of the material becomes high enough, degeneracy pressure is overcome and a thermonuclear runaway begins.

Nuclear energy released during the nova outburst causes an expansion of the WD atmosphere and the loss of mass from the system through a variety of mechanisms. One
is ``reverse" RLOF that occurs when the atmosphere of the outbursting WD reaches its RL radius of $\lesssim R_{\odot}$\cite{Macdonald80,Shara+86,Livio+90,Shankar+91,Lloyd+97}. 
The WD can also lose mass during the nova outburst through optically-thick winds\cite{Bath&Shaviv76,Kato&Hachisu94} driven by the momentum imparted by locally near- or super-Eddington luminosities. The latter may occur near the iron opacity bump in the envelope at temperatures of $T \approx 10^{5.2}$ K\cite{Kato&Hachisu94}; however, the dynamics of this region of the envelope is complicated by magneto-hydrodynamical instabilities\cite{Jiang+15}, making it challenging to predict the properties of the outflow such as its mass loss rate and velocity.

Both RLOF and super-Eddington winds can occur at different stages of the nova outburst. 
There is observational evidence, from multi wavelength light curves\cite{Chomiuk+14a,Nelson+2014}, optical\cite{Schaefer+14} and radio imaging\cite{Chomiuk+14}, as well as the evolution of optical spectral lines\cite{McLaughlin57,Friedjung+63,Williams&Mason10,Ribeiro+13}, for at least two distinct phases of mass loss during the nova outburst (see Supplementary Figure~1 for a schematic illustration). 
Chomiuk et al.\cite{Chomiuk+14} argue that at early times, the nova outburst produces a slow ejection of
mass concentrated in the equatorial plane of the binary. One may tentatively associate this period with a RLOF or ``common envelope" phase in the nova evolution\cite{1989ApJ...339..268S}. Alternatively, the outflow could be intrinsically focused in the equatorial plane for reasons unrelated to the binary companion, such as the rotation of the white dwarf.\cite{Lloyd+97} At later times a faster wind emerges, likely from from deeper within the potential well of the WD and with a more spherical geometry. The collision between these fast and slow components in the equatorial plane provides a natural location for the production of strong shocks and their accompanying multi-wavelength emission.\cite{Chomiuk+14} The fast component continues to expand freely along the polar direction, imparting a bipolar morphology to the ejecta\cite{Soker+89,Lloyd+93,Chomiuk+14,Metzger+14}. 

\section{Phase I of ASASSN-16ma: Slow RLOF Outflow Phase}
\label{sec:slow}

The light curve of \nova\ begins during Phase I with a slow rise over about 2 weeks (Figure~1). We attribute the emission during this early phase to a slow outflow caused by RLOF from the binary of mass loss rate $\dot{M}_{\rm s}$ and velocity $v_{\rm s} \lesssim 400$ km s$^{-1}$. The latter is comparable to the orbital velocity/escape speed at the orbit of the main sequence companion of $(G[M_{\rm wd}+M_{\star}]/2a)^{1/2} \approx 400$ km s$^{-1}$. 

Observationally, the H$\alpha$ emission at MJD 57688.5 (day 2) has a two-component structure\cite{Luckas16}. Most of the flux is in a narrow component that is unresolved at the resolution of the spectrum (FWHM 520 km s$^{-1}$) and thus broadly consistent with the velocity of the slow outflow that we assume (Supplementary Figure~2). A few percent of the flux is in a broad component that extends to velocities of $\sim 1100$ km s$^{-1}$, though the FWHM of this component cannot be measured precisely. This early high-velocity material could represent a low-mass shell ejected by the dynamical phase of the thermonuclear runaway that is not central to the overall energsetics of the nova.

The slow, equatorially-focused mass loss could result from frictional heating by the binary inside a common envelope or could originate from an anisotropic wind launched from a rotating WD\cite{Macdonald80,Shara+86,Livio+90,Kato&Hachisu91,Shankar+91,Lloyd+97}. Binary mass loss can occur from the outer L2 Lagrange point (that furthest from the WD), naturally resulting in an equatorially-concentrated outflow through a sprinkler-like spiral ejection\cite{Shu&Lubow79,Pejcha+16a,Pejcha+16b}. For a binary of mass ratio $q = M_{\star}/M_{\rm wd} \approx 0.5$, Pejcha et al.\cite{Pejcha+16a} find that mass loss from L2 of cold material corotating with the binary experiences torques, but reaches asymptotic velocities of only $\sim 100$ km s$^{-1}$. Hotter matter, as might be expected due to radioactive decay\cite{Starrfield+98} or frictional heating by the binary, can reach higher asymptotic velocities, approaching or moderately exceeding the binary escape speed\cite{Livio+90,Pejcha+16b}. 

In conventional nova models, the optical/UV emission is powered by the outwards diffusion of thermal energy generated by nuclear burning near the WD surface\cite{Wolf+13,Hillman+14}. 
However, another possible source of optical emission is that powered by internal shocks within the RLOF outflow. Pejcha et al.\cite{Pejcha+16a} shows that spiral streams which occur following mass loss from L2 undergso internal shocks at a radial distance of $r_{\rm sh} \approx 8a \approx 10^{12}$ cm (Supplementary Figure~1, left panel). Reprocessed thermal energy from these shocks powers a luminosity given by
\be
L_{\rm L2} \approx \frac{1}{2}\dot{M}_{\rm s}(\Delta v)^{2} \approx 1.7\times 10^{37}\,{\rm ergs\,s^{-1}}\left(\frac{\dot{M}_{\rm s}}{10^{-4}M_{\odot}\,{\rm week^{-1}}}\right)\left(\frac{v_{\rm s}}{400\,{\rm km\,s^{-1}}}\right)^{2}, 
\ee
where $\Delta v \approx v_{\rm s}/4$ is the velocity dispersion between the spiral arms\cite{Pejcha+16a}. The effective temperature of this emission is approximately 
\be
T_{\rm eff} = \left(\frac{L_{\rm L2}}{2\pi \sigma r_{\rm sh}^{2}}\right)^{1/4} \approx 2\times 10^{4}{\rm K}\,\left(\frac{L_{\rm L2}}{10^{38}\,{\rm ergs\,s^{-1}}}\right)^{1/4},
\ee 
 where $2\pi r_{\rm sh}^{2}$ is the approximate surface area of the emitting ``plates" produced by the colliding spiral arms. This emission will peak in the optical/UV for typical parameters.
 The equatorial wind ejecta is opaque in the continuum at this phase leading to spectra that, despite the somewhat unusual geometry, could well appear consistent with the usual optically-thick early stage of the nova outburst. L2 spiral shocks powering early-time optical emission from novae would be consistent with observations of periodicities at the orbital period in the early-time light curves of some novae\cite{2010AJ....140...34S,Mroz+16,Finzell+17}. 

Equating $L_{\rm L2}$ to the optical luminosity during Phase I, we find that for $v_{\rm s} = 400$ km s$^{-1}$ the total emission can be explained by the ejection of a total mass of $M_{\rm s} \approx 3\times 10^{-4}M_{\odot}$ over two weeks, with the mass loss rate reaching values approaching $\dot{M}_{\rm s} \sim 10^{-3}M_{\odot}$ week$^{-1}$ during the final few days of Phase I. Lower ejecta masses are needed if the wind velocity is higher than 400 km s$^{-1}$.

\subsection{Opacity of Slow Outflow.}

The slow outflow not only provides a potential source of luminosity during early phases in the outburst, it also lays the groundwork for the observed emission in later phases by providing the medium in which the faster wind will produce a shock ($\S\ref{sec:fast}$). The slow outflow acts as a barrier between the shocks and an external observer, although the effects of the barrier may change with viewing angle.

We assume that the slow outflow results in a torus-like structure that subtends a solid angle $\Omega$ (Supplementary Figure~1). Making the simplifying assumption that the outflow velocity is constant during Phase I, the radial density profile in the slow outflow at radius $r \approx v_{\rm s}t'$ (where time $t'$ is measured prior to the end of slow wind phase) is given by 
\be
n_{\rm s} = \frac{\dot{M}_{\rm s}(r)}{4\pi f_{\Omega} v_{\rm s} r^{2}\mu m_p},
\ee 
where $\mu = 0.74$ is the mean molecular weight of fully ionized gas with a composition typical of that of classical novae\cite{Schwarz+07} (we assume abundances as follows: [He/H] = 0.08, [N/H] = 1.7, [O/H] = 1.3, [Ne/H] = 1.0, [Mg/H] = 0.7, [Fe/H] = 0.7), and $f_{\Omega} \approx 0.3$ is the assumed fraction of the total solid angle subtended by the slow outflow. The mass column external to radius $r$ is then given by
\be
\Sigma = \mu m_p\int n dr = \frac{\dot{M}_{\rm s}}{4\pi f_{\Omega} v_{\rm s} r }
\ee
This column determines the optical depth of the torus in three important frequency ranges: optical, X-rays, and GeV $\gamma$-rays.

At optical frequencies, the continuum opacity of the expanding torus is dominated by electron scattering and Doppler-broadened Fe lines\cite{Pinto&Eastman00}, with a characteristic value $\kappa_{\rm opt} \approx 0.01-0.3$ cm$^{2}$ g$^{-1}$, such that the optical depth external to radius $r$ is 
\be
\tau_{\rm opt} = \Sigma \kappa_{\rm opt} \approx 22\left(\frac{\kappa_{\rm opt}}{0.1\,\rm cm^{2}\,g^{-1}}\right)\left(\frac{\dot{M}_{\rm s}}{10^{-4}M_{\odot}\,{\rm week}^{-1}}\right)\left(\frac{r}{10^{13}\,{\rm cm}}\right)^{-1}\left(\frac{v_{\rm s}}{400\,{\rm km\,s^{-1}}}\right)^{-1}.
\label{eq:tauopt}
\ee
At X-ray frequencies, the bound-free opacity of neutral gas for the assumed ejecta composition is approximately given by\cite{Vlasov+16}
\be
\kappa_X \approx 2000\left(\frac{E_{X}}{\rm keV}\right)^{-2}\,{\rm cm^{2}\,g^{-1}},
\ee
where $E_{\rm X}$ is the X-ray energy, resulting in an X-ray optical depth of
\be
\tau_X = \Sigma \kappa_X \approx 4\times 10^{5}\left(\frac{\dot{M}_{\rm s}}{10^{-4}M_{\odot}\,{\rm week}^{-1}}\right)\left(\frac{r}{10^{13}\,{\rm cm}}\right)^{-1}\left(\frac{v_{\rm s}}{400\,{\rm km\,s^{-1}}}\right)^{-1}\left(\frac{E_{X}}{\rm keV}\right)^{-2}.
\label{eq:tauX}
\ee
The very high value of $\tau_X$ implies that at keV energsies detectable by {\it Swift} XRT, the slow wind is opaque to X-rays at radii of interest for several weeks or longer after their onset (see below). 
Thus, even if the shock power is as high as $L_{\rm X} \approx 10^{39}$ ergs s$^{-1}$ (see below), we can have an observed X-ray luminosity of $L_{\rm X,obs} = L_{\rm X}\exp(-\tau_X) \lesssim 10^{33}$ ergs s$^{-1}$, below the XRT upper limit (see Methods). Some novae appear as hard thermal X-ray sources at later times (e.g.~Ref.\cite{Mukai&Ishida01}), presumably once the column of gas ahead of the shocks is significantly lower. One can also have $\tau_X \lesssim 1$ at higher energsies $\gtrsim 10$ keV (e.g.~in the NuSTAR band), making it possible to detect very hard X-rays if the shocks are high velocity (high temperature) and/or if there is non-thermal emission extending down from the $\gamma$-ray band\cite{Vurm&Metzger16}. 

The opacity of gas to high energy $\gamma$-rays of energy $x = E_{\gamma}/m_e c^{2} \gg 1$ is due to inelastic electron scattering in the Klein-Nishina regime, for which the opacity is approximately given by\cite{Klein&Nishina29}
\be
\kappa_{\rm KN} \approx \frac{\sigma_T}{m_p}\frac{3}{8}\frac{1}{x}\left({\rm ln}[2x]+\frac{1}{2}\right) \underset{x \approx 600}\approx 0.0019\left(\frac{E_{\gamma}}{300\,{\rm MeV}}\right)^{-1}\,{\rm cm^{2}\,g^{-1}}
\ee
and thus the optical depth of the slow wind in the LAT band is given by
\be
\tau_{\gamma} = 0.42\left(\frac{\dot{M}_{\rm s}}{10^{-4}M_{\odot}\,{\rm week}^{-1}}\right)\left(\frac{r}{10^{13}\,{\rm cm}}\right)^{-1}\left(\frac{v_{\rm s}}{400\,{\rm km\,s^{-1}}}\right)^{-1}\left(\frac{E_{\gamma}}{300\,{\rm MeV}}\right)^{-1}.
\label{eq:tauKN}
\ee
Thus, at the base of the outflow near the binary separation $r \sim 10^{11}$ cm the outflow can be opaque to $\sim 0.1-1$ GeV $\gamma$-rays, an important below in describing the delayed onset of the $\gamma$-ray emission.

At energies $E_{\gamma} \gg 2 m_e c^{2}$, $\gamma$-rays can also be attenuated by pair creation processes associated with their interaction with the electric field of ions in the ejecta. At energies well above the pair creation threshold $E_{\gamma} \gg 2 m_e c^{2}$, the opacity due to this process is approximately given by $\kappa_{\gamma -n} \approx \alpha_{\rm fs}(Z^{2}/Am_{\rm p})\sigma_{T} \approx 0.0028(Z^{2}/A)$ cm$^{2}$ g$^{-1}$ and thus can be comparable to the Klein-Nishina cross section in the LAT energy range.

\section{Phase II/III: Collision Between Fast Wind and Slow Torus}
\label{sec:fast}

{\bf Fast Wind.} Starting near the end of Phase I, we hypothesize that the nova outflow undergsoes an abrupt transition from slow equatorial mass loss of velocity $v_{\rm s} \lesssim 1000$ km s$^{-1}$ to a faster outflow with a velocity of $v_{\rm f} \gtrsim 2000$ km s$^{-1}$, more representative of the escape speed closer to the WD surface and what might be expected from super-Eddington winds. This behavior is consistent with the evolution of the H$\alpha$ velocity profile in ASASSN-16ma, which grows from a half width at zero intensity (HWZI) velocity of $\lesssim 1100$ km s$^{-1}$ around MJD 57692 (Day 6) to a HWZI of 2200 km s$^{-1}$ by MJD 57704 (Day 18) into Phase III, when the $\gamma$-ray emission is observed (Supplementary Figure~2). Such transitions in nova outbursts from slow or nearly ``hydrostatic" to faster wind states have been discussed extensively in the literature\cite{Kato&Hachisu11} and the connection of light curve changes to spectroscopic changes is well documented\cite{Gallagher&Starrfield78,Kato+02,Csak+05,Tanaka+11}.

Unaffected by the gravity of the companion star (since $v_{\rm f} \gg v_{\rm esc}$), we assume that this faster wind will possess a more spherical geometry than the slower early-time flow, as illustrated schematically in Supplementary Figure~1. The density of the fast wind at radius $r$ from the binary for a steady mass loss rate $\dot{M}_{\rm f}$ is then approximated as
\be
n_{\rm f} \approx \frac{\dot{M}_{\rm f}}{4\pi v_{\rm f} r^{2}\mu m_p},
\ee
where a characteristic value of $v_{\rm f} \approx 2000$ km s$^{-1}$ is motivated by the approximate H$\alpha$ line width at late times (Supplementary Figure~2).

\subsection{Shock Dynamics. }
The fast wind developed near the beginning of Phase II collides with the previous slow wind emitted during Phase I, driving a forward shock-reverse shock structure outwards\cite{Metzger+14}. 
As we now argue, these shocks power the observed $\gamma$-ray emission and, contrary to standard models, the majority of the optical emission. 

We assume that the shocks are radiative, as we will verify below. Cold gas behind the radiative shock collects in a cold and clumpy shell of mass $M_{\rm sh}$ and velocity $v_{\rm sh}$, the latter identical to the velocity of both the forward and reverse shocks in the rest frame of the WD\cite{Metzger+14}. After the onset of the fast outflow ($t = 0$), the mass of the cold shell grows as\cite{Metzger+14}
\be
\frac{dM_{\rm sh}}{dt} = f_{\Omega}\dot{M}_{\rm f}\left(\frac{v_{\rm f} - v_{\rm sh}}{v_{\rm f}}\right) + \dot{M}_{\rm s}\left(\frac{v_{\rm sh}-v_s}{v_s}\right), \ee
while its momentum grows as
\be
\frac{d}{dt}\left(M_{\rm sh}v_{\rm sh}\right) = f_{\Omega}\dot{M_{\rm f}}(v_{\rm f}-v_{\rm sh}) + \dot{M}_{\rm s}(v_{\rm sh}-v_{\rm s}),
\ee
where $\dot{M}_{\rm s}$ is evaluated at radius $r$ according to its value a time $t \approx r/v_{\rm s}$ before the onset of the fast wind. 
For simplicity we assume both that $\dot{M}_{\rm s}$ was constant in time prior to the onset of the fast outflow, while $\dot{M}_{\rm f}$ is constant after that point. 

A numerical integration of the above equations shows that an approximate steady state is quickly reached ($dv_{\rm sh}/dt \approx 0$), in which the shell gains most of its momentum from the fast wind and most of its mass by sweeping up the slow ejecta shell. In this limit that $v_{\rm f} \gg v_{\rm s}$ and $\dot{M}_{\rm f} \lesssim \dot{M}_{\rm s}f_{\Omega}^{-1}$ we have
\be
\frac{v_{\rm sh}}{v_{\rm s}} \approx \left(\frac{\dot{M}_{\rm f}v_{\rm f}f_{\Omega}}{\dot{M}_{\rm s}v_{\rm s}}\right)^{1/2} \equiv \xi.
\ee
The cold shell radius thus grows approximately as 
\be
R_{\rm sh} \approx v_{\rm sh}t \approx 3.5\times 10^{12}\,\,\,\xi \left(\frac{v_{\rm s}}{400\,{\rm km\,s^{-1}}}\right)\left(\frac{t}{1\rm \,day}\right)\,{\rm cm},
\label{eq:Rsh}
\ee 
where $\xi \sim 1-3$ for typical parameters. In detail, the shell velocity will accelerate as it propagates to larger radii as the fast wind gains in power and velocity with time. It is thus tempting to associate the low velocity component of the H$\alpha$ velocity profile in ASASSN-16ma (Supplementary Figure~2), which grows in width and develops a P Cygni profile indicative of a dense shell near the end of Phase I, with this cold swept-up shell instead of, or in addition to, the unshocked slow component. 

The power dissipated at the reverse ($L_r$) and forward ($L_f$) shocks are given by
\be
L_r = \frac{9}{32}f_{\Omega}\frac{\dot{M}_{\rm f}}{v_{\rm f}}(v_{\rm f}-v_{\rm sh})^{3}
\ee
and
\be
L_f = \frac{9}{32}\frac{\dot{M}_{\rm s}}{v_{\rm s}}(v_{\rm sh}-v_{\rm s})^{3},
\ee
respectively. If $v_{\rm sh} \ll v_{\rm f}$ then the total shock power $L_{\rm sh}$ is dominated by the reverse shock,
\be
L_{\rm sh} \approx L_r \approx \frac{9}{32}f_{\Omega}\dot{M}_{\rm f}v_{\rm f}^{2} \approx 1.0\times 10^{39}\,{\rm ergs\,s^{-1}}\left(\frac{\dot{M}_{\rm f}}{10^{-4}M_{\odot}\,{\rm week^{-1}}}\right)\left(\frac{v_{\rm f}}{2000\,{\rm km\,s^{-1}}}\right)^{2},
\label{eq:Lsh}
\ee
notably comparable to the bolometric output of the nova. The shock is radiative if the radiative cooling time $t_{\rm cool}$ is shorter than the expansion time $t_{\rm exp}$. For free-free cooling, their ratio is given by 
\begin{eqnarray}
&&\frac{t_{\rm cool}}{t_{\rm exp}} = \frac{3 k T_{\rm sh}}{8\mu \Lambda n_{\rm f}(R_{\rm sh})}
\frac{v_{\rm sh}}{R_{\rm sh}} \approx
\nonumber \\
&3\times 10^{-4}&\xi^{2}\left(\frac{\dot{M}_{\rm f}}{10^{-4}M_{\odot}\,{\rm week^{-1}}}\right)^{-1}\left(\frac{v_{\rm f}}{2000\,{\rm km\,s^{-1}}}\right)^{2}\left(\frac{v_{\rm s}}{400\,{\rm km\,s^{-1}}}\right)^{2}\left(\frac{t}{\rm 1\,day}\right) \nonumber \\
\end{eqnarray}
where $\Lambda \approx 3\times 10^{-27}T_{\rm sh}^{1/2}\,{\rm ergs\,cm^{3}\,s^{-1}}$ is the free-free cooling rate for a shock of temperature $T_{\rm sh} = 6.7\times 10^{7}(v_{\rm f}/3000\,{\rm km\,s^{-1}})^{2}$ K (this is the initial temperature before cooling). The shocks are thus radiative ($t_{\rm cool} \ll t_{\rm exp}$) on timescales of at least a couple weeks for our fiducial parameters, a result with several key implications we now describe.

\subsection{Optical Emission. }
\label{sec:optical}

Gas passing through a radiative shock cools and loses pressure support over a distance $t_{\rm cool}v_{\rm sh}/4$, much smaller than the radius of the shock $R_{\rm sh}$. A defining feature of radiative shocks is that they radiate the dissipated energy with nearly 100\% efficiency, as opposed to adiabatic shocks ($t_{\rm cool} \gtrsim t_{\rm exp}$), for which most of the post-shock thermal energy is instead lost to adiabatic expansion. 

Radiative shocks in novae produce thermal X-rays of luminosity comparable to the total shock power $L_X \approx L_{\rm sh}$ and temperature $kT_{\rm sh} \approx 5.8(v_{\rm f}/2000{\rm\, km\,s^{-1}})^{2}$ keV. However, the optical depth of the ejecta at X-ray frequencies $E_{\rm X} = kT_{\rm sh}$ at the shock radius $r = R_{\rm sh}$ (eq.~\ref{eq:Rsh}) is very large, (eq.~\ref{eq:tauX})
\be
\tau_X(R_{\rm sh}) \approx 5\times 10^{3}\xi^{-1}\left(\frac{t}{\rm 1\,week}\right)^{-1}\left(\frac{\dot{M}_{\rm s}}{10^{-4}M_{\odot}\,{\rm week}^{-1}}\right)\left(\frac{v_{\rm s}}{400\,{\rm km\,s^{-2}}}\right)^{-2}\left(\frac{v_{\rm f}}{\rm 2000\,{\rm km\,s^{-1}}}\right)^{-4},
\label{eq:tauX2}
\ee
Because $\tau_X(R_{\rm sh}) \gg 1$, the X-ray luminosity is absorbed and reprocessed through line and continuum emission processes to optical/UV frequencies, where the much lower opacity more readily allows radiation to escape\cite{Roth+16}. In other words, the shock will ultimately produce an optical/UV luminosity of
\be
L_{\rm opt} \approx L_{\rm X} \approx L_{\rm sh} \sim 10^{38}-10^{39}\,{\rm ergs\,s^{-1}},
\ee
as observed.

As we discuss in the next section, the $\gamma$-ray emission is powered by relativistic particles accelerated at the shock. The observed correlation between optical and $\gamma$-ray light curves over the first $\sim$ week of the Phase III light curve therefore strongly suggests that a large fraction of the optical emission is shock-powered. This motivates our fiducial values for $v_{\rm f}$ and $\dot{M}_{\rm f}$, as these are the values required to match the shock luminosity from eq.~\ref{eq:Lsh} to the observed peak optical luminosity of $\sim 10^{39}$ ergs s$^{-1}$ at our fiducial distance. 

That shocks power the peak emission in some novae contrasts with previous models which instead attribute the emission to diffusion through the hydrostatic white dwarf atmosphere\cite{Hillman+14}. This new mechanism also provides a possible answer to the mystery of how classical novae---often modeled as hydrostatic phenomena---can produce luminosities well in excess of the Eddington luminosity of $L_{\rm edd} \approx 1.5\times 10^{38}$ ergs s$^{-1}$ for a solar mass object\cite{Chochol+93,Schwarz+98,Schwarz+01}. Some previous models have invoked ``porosity" in the radiation-dominated envelope to reduce the opacity and hence increase the {\it effective} Eddington luminosity of the WD\cite{Shaviv01,Kato&Hachisu07}. However, it is unclear whether this effect can explain luminosities $\sim 10^{39}$ ergs s$^{-1} \sim 10L_{\rm edd}$ inferred for some novae, such as Nova LMC 1991\cite{Schwarz+01} and potentially ASASSN-16ma (keeping in mind that the distance to ASASSN-16ma is uncertain). By contrast, shock-powered emission is not bound by the Eddington luminosity provided that the kinetic power of the fast outflow is itself super-Eddington\cite{Quataert+16}. 

Gas passing through a radiative shock will collect into a thin shell with a density $\sim 10^{3}-10^{4}$ times higher than the upstream medium. This gas is subject to thermal and geometrical thin-shell instabilities\cite{Chevalier&Imamura82,Vishniac83}, which could potentially contribute to the ubiquitous clumpy geometry inferred for nova ejecta\cite{Williams92,Saizar&Ferland94}.

As discussed below, a final consequence of the discovery that the peak emission in \nova\ is shock-powered is that the ratio of the optical to $\gamma$-ray luminosities can be used to directly constrain the particle acceleration efficiency\cite{Metzger+15,Vurm&Metzger16}.

Although we have proposed shock-powered origins for Phases I and III of ASASSN-16ma's light curve, we have not yet addressed Phase II. Phase II could represent a gradual transition phase between the fast and slow outflows, during which the wind velocity and mass loss rate are in some sense intermediate. Alternatively, we propose a scenario in which the fast outflow begins near the end of Phase I, with the gradual rise of the optical light curve during Phase II instead being caused by the finite time required for the shock-powered emission to diffuse out through the slow outflow. 

Optical radiation can only escape from the shock if the photon diffusion timescale through the slow outflow, $\sim \tau_{\rm opt}(R_{\rm sh}/c)$, is short compared to the outflow expansion rate, $t_{\rm exp}$. Equating these timescales gives the minimum optical rise time,
\be
t_{\rm rise} \approx \frac{\kappa_{\rm opt}\dot{M}_{\rm s}}{4\pi f_\Omega c v_{\rm s}} \approx 0.08\,{\rm day}\left(\frac{\kappa_{\rm opt}}{0.1\,\rm cm^{2}\,g^{-1}}\right)\left(\frac{\dot{M}_{\rm s}}{10^{-4}M_{\odot}\,{\rm week}^{-1}}\right)\left(\frac{v_{\rm s}}{400\,{\rm km\,s^{-1}}}\right)^{-1}.
\label{eq:trise}
\ee
For the values of $\dot{M}_{\rm s} \sim 10^{-3}M_{\odot}\,{\rm week}^{-1}$ needed to explain the optical luminosity near the end of Phase I, the value of $t_{\rm rise}$ is indeed comparable to the duration of Phase II. In this picture, at times $t \ll t_{\rm rise}$ (the earliest parts of Phase II), the fast wind and shocks have already commenced, but the observed optical emission is less than the true luminosity at the shock due to adiabatic losses as the photons diffuse through the radially expanding ejecta. By contrast, at times $t \gg t_{\rm rise}$, the optical emission will more faithfully track the shock power and thus the mass loss history from Phase I in reverse, as long as the shock remains radiative. 

Although the ejecta is transparent to $\gamma$-rays (except possibly at early times; see below), the reprocessing picture we have described may imprint a moderate viewing angle dependence on the optical emission. X-rays are absorbed and reprocessed very close to the shock, and the optical emission will diffuse outwards from that point to the observer. Given the geometry we propose (Supplementary Figure~1), optical radiation might diffuse first from the shocks into the lower density polar regions and then radially outwards from that point. This could imprint a moderate viewing angle dependence on the optical emission, insofar that events viewed at higher latitudes would be moderately optically brighter than those viewed in the equatorial plane for the same shock power ($\gamma$-ray luminosity). Given that the ratio of $L_{\gamma}/L_{\rm opt}$ is the lowest for V339 Del (Supplementary Figure~3), this would be consistent with the relatively low inclination of 45$\pm 10$ degrees inferred for this system\cite{Shore+16}. Likewise, the major dust extinction event observed in V1324 Sco\cite{Finzell+17} may suggest an equatorial viewing angle\cite{Derdzinski+16} consistent with its higher value of $L_{\gamma}/L_{\rm opt}$ inferred for this event. Taking this interpretation literally, the intermediate value of $L_{\gamma}/L_{\rm opt}$ for~\nova~would suggest a relatively high inclination angle for this system intermediate between that of V1324 Sco and V339 Del.

If optical emission in novae is powered by shocks, then another prediction is that we might generically expect detectable $\gamma$-ray emission from classical novae with strong secondary peaks in the optical light curve (e.g.~Type C defined in Ref.\cite{Strope+10}). More broadly, we encourage future searches for $\gamma$-ray emission from a larger sample of classical novae, specifically focusing on epochs in which the optical flux is highest. 

\section{Gamma-Ray Spectral Fitting}

The continuum $\gtrsim 100$ MeV $\gamma$-ray emission observed in novae by Fermi LAT is the result of relativistic protons or electrons accelerated at the shocks\cite{Martin&Dubus13,Ackermann+14,Metzger+15,Metzger+16,Vurm&Metzger16}, probably due to diffusive particle acceleration\cite{Caprioli&Spitkovsky14a}. In the leptonic scenario, the $\gamma$-ray emission is the result of direct electron acceleration, with the dominant non-thermal radiative processes being relativistic bremsstrahlung and inverse Compton (IC) scattering off the background nova optical light. In hadronic models, when the energy in accelerated protons dominates over electrons, the $\gamma$-rays are instead mainly produced by the production and decay of neutral pions ($\pi_0$), created by proton-proton collisions. These collisions also produce charged pions ($\pi_{\pm}$) that ultimately decay into relativistic electron-positron pairs carrying energy comparable to that from $\pi_{0}$ decay. 

We model the Fermi LAT SED from \nova\ using the emission model developed by Vurm \& Metzger\cite{Vurm&Metzger16}. This model accounts for IC emission from leptons using a seed radiation field estimated directly from the observed optical luminosity, relativistic bremsstrahlung emission from electron and positrons interacting with the ambient gas, and synchrotron emission, calculated assuming that a fraction $\epsilon_B$ of the shock-dissipated energy is placed into the magnetic field $B$, i.e. $B^{2}/8\pi = (3/4)\epsilon_{B}\rho v_{\rm sh}^{2}$. We account for the effects of compression of the thermal gas on the relativistic particle distribution and its resulting non-thermal emission. This is important because the cooling timescale of the thermal plasma can be shorter than the non-thermal cooling timescale for the range of particle Lorentz factors contributing to the LAT emission.

\subsection{Leptonic Model}

The theoretical $\gamma$-ray spectrum in the leptonic model is shown in Figure~2, along with the Fermi-LAT data. The relativistic electrons are injected at the shock with a power-law distribution with equal energy per logarithmic $\gamma$ interval. Nevertheless, there is significant spectral curvature in the observed LAT spectrum above 300 MeV. In the leptonic model fit, the spectral shape is determined by the sum of IC and bremsstrahlung emission, which have comparable magnitudes at the required parameters. The turnover below $\sim 300$~MeV is due to Coulomb losses, whereby the relativistic electrons rapidly share their energy with the thermal population of particles without producing radiation. In light of the above, care should be taken when inferring the energy distribution of accelerated particles from the observed spectrum.

Despite the relatively good fit, we argue that leptonic models are disfavored for the following reasons:
\begin{enumerate}
\item
Acceptable fits are found only for low post-shock magnetization, $\epsilon_{\rm B} \lesssim 10^{-6}$. For stronger magnetic fields, synchrotron emission saps energy from the relativistic electrons that would otherwise produce $\gamma$-ray emission (the synchrotron emission itself occurs at much lower photon energsies, well below the LAT bandpass). This steepens the spectrum beyond what is allowed by the data and results in an energy crisis. However, as discussed below in $\S\ref{sec:Bfield}$, stronger magnetic fields $\epsilon_{\rm B} \gtrsim 10^{-4}$ are needed to accelerate particles to sufficiently high energsies to explain the highest energy $\gamma$-rays observed. 
\item
A high optical depth for the shocks $\tau_{\rm opt}$ during the first week after the onset of the fast wind (see eq.~\ref{eq:tauopt}) enhances the optical radiation density as $u_{\rm opt} = (1+\tau_{\rm opt}) L_{\rm opt}/(4\pi r^2 c)$; IC cooling of relativistic electrons is correspondingly enhanced, which softens the spectrum below $\sim 300$~MeV. Acceptable fits could only be found for $\tau_{\rm opt} \lesssim$~a~few (note that $\tau_{\rm opt}\ll 1$ is assumed in Figure~2), much smaller than our estimate for the optical depth of the slow outflow in eq.~\ref{eq:tauopt}. 
Such a low optical depth for the shocks is incompatible with our explanation for the sudden onset of the $\gamma$-ray emission (Supplementary Figure~4). 
\item
The required electron acceleration efficiency $\epsilon_{\rm e}\sim 2.5\times 10^{-3}$ is greater than the efficiencies of $\sim 10^{-4}$ inferred from some observations of supernova remnants\cite{Morlino&Caprioli12} as well as particle-in-cell plasma simulations of shock acceleration\cite{Kato15,Park+15}. However, we note that recent work modeling supernova remnants in M33\cite{2017MNRAS.464.2326S} has found higher values of $\epsilon_{\rm e} (> 10^{-3}$). Higher values are also inferred from radio continuum emission of the Galaxy and other star-forming galaxies from the FIR-radio correlation (e.g.~Ref.\cite{Lacki+10,Lacki&Thompson10}), so this argument disfavoring leptonic models is somewhat weaker than those above.

\end{enumerate}
 
 \subsection{Hadronic Model}
 
The LAT $\gamma$ rays in the hadronic model are produced by $\pi_0$ decay (Figure~2).
In contrast, IC and bremsstrahlung emission from relativistic pairs produced by $\pi_{\pm}$ decay
make a negligible contribution in the LAT band, due to strong synchrotron losses in the compression-enhanced magnetic field.
Overall, the hadronic model is more robust compared to the leptonic scenario;
the $\gamma$-ray emission depends mainly on the acceleration parameters, but weakly on other quantities such as the ambient radiation density and shock magnetization.

Notably, assuming that the optical emission is entirely shock-powered,
this enables us to estimate the non-thermal particle acceleration efficiency\cite{Metzger+15} to be $\epsilon_p \approx 5 \times 10^{-3}$. An ion acceleration efficiency of $\sim 1$\% is to be compared to the maximum acceleration efficiency of 20\% inferred by hybrid particle-in-cell simulations of non-relativistic shocks\cite{Caprioli&Spitkovsky14a} for cases when the upstream magnetic field is aligned parallel to the shock normal. The low inferred acceleration efficiency is perhaps not unexpected since we in general expect the magnetic field in the fast wind to be oriented primarily in the azimuthal direction (i.e., perpendicular to the radial direction) and hence the probable shock normal, since the geometry is driven by radial expansion. One prediction of the hadronic model is the simultaneous emission of $\gtrsim 10-100$ GeV neutrinos from pion decay; however, this signal would only be detectable by IceCube for particularly nearby novae.\cite{Razzaque+10,Metzger+16}

This general technique for constraining particle acceleration relies on the fact that the reprocessed optical emission from a radiative shock constrains the {\it total} (and energetically dominant) thermal power of the shock, while the $\gamma$-ray emission gives us the total {\it non-thermal} power since the nova ejecta is sufficiently dense to render it an effective calorimeter for relativistic particles\cite{Metzger+15}. As shown in Supplementary Figure~3, the ratio of $\gamma$-ray to optical luminosities for the other LAT-detected novae are all within an order of magnitude of that of \nova. Thus, assuming that all events accelerate particles at the same $\sim$ 1\% efficiency inferred for \nova~also implies that much of the optical emission in these other events is also shock powered\cite{Metzger+15}.

\section{Magnetic Field Amplification}
\label{sec:Bfield}

In order to explain $\gamma$-ray emission extending up to photon energsies $\epsilon_{\gamma,\rm max} \gtrsim 10$ GeV, particles must be accelerated up to maximum energsies $E_{\rm max} \gtrsim 10\epsilon_{\gamma,\rm max} \sim 100$ GeV. Metzger et al.\cite{Metzger+16} show that the maximum particle energy, accounting for the confinement of particle acceleration region to the thin photo-ionized layer ahead of the shock, is given by (their eq.~23)
\begin{eqnarray}
E_{\rm max} \approx 19\,\,{\rm GeV}\,\,\xi \left(\frac{\epsilon_{B}}{10^{-4}}\right)^{1/2}\left(\frac{v_{\rm f}}{2000\,{\rm km\,s^{-1}}}\right)^{7/2}\left(\frac{v_{\rm s}}{400\,{\rm km\,s^{-1}}}\right)\left(\frac{t}{\rm 1\, day}\right) \left(\frac{\dot{M}_{\rm f}}{10^{-4}M_{\odot}\,{\rm week^{-1}}}\right)^{-1/2},
\label{eq:emax}
\end{eqnarray}
where we have focused on the reverse shock by taking as the upstream density that of the fast wind, $n_{\rm f} = \dot{M}_{\rm f}/(4\pi R_{\rm sh}^{2}v_{\rm f}\mu m_p)$, and $t$ is again measured with respect to the onset of the fast wind. 

From this we conclude that magnetic equipartition factors of $\epsilon_{B} \gtrsim 10^{-4}$ are required to achieve the required values of $E_{\rm max} \gtrsim 100$ GeV on timescales of $\gtrsim 1$ day into Phase 2. 
Such high values of the magnetic field cannot be explained as primordial fields carried from the white dwarf surface. To see this, note that the fast wind must be launched from radii $R_{\rm in} \lesssim 0.01-0.1R_{\odot} \lesssim 10^{-4}-10^{-3}R_{\rm sh}$ given its high velocity, much smaller than the shock radius $R_{\rm sh} \gtrsim 10^{12}$ cm (eq.~\ref{eq:Rsh}). Even under the optimistic assumption that the geometry of the magnetic field is purely toroidal, and hence is diluted by lateral expansion with radius as $B \propto 1/r$, the required surface magnetic field would need to be $\sim 10^{3}-10^{4}$ times larger at the WD surface than at the shock. For $\epsilon_B \gtrsim 10^{-4}$ at the shock, this would require a highly super-equipartition field near the WD surface $\epsilon_{B} \sim 1$, which is clearly unphysical. Instead, the $\gamma$-ray emission from novae such as \nova\ provides strong evidence for magnetic field {\it amplification} at non-relativistic shocks. The evidence for such amplification has been hotly debated in other astrophysical shocks, such as those with similar velocities in supernova remnants\cite{Berezhko+03,Ressler+14}. 

Magnetic field amplication also implies that the shocks will produce radio synchrotron emission, from leptons accelerated directly at the shock and from the secondary electron/positrons pairs produced by $\pi^{\pm}$ decay.\cite{Murase+11,Murase+14,Vlasov+16} At times of interest in this work, centimeter wavelength emission will be highly attenuated due to free-free absorption by the large column of ionized gas ahead of the shocks.\cite{Metzger+14} However, synchrotron emission becomes visible on timescales of months (e.g.~Ref.\cite{Weston+16}), once the shocks propagate to lower densities and the level of ionizing radiation from the shocks subsides.\cite{Vlasov+16}

\section{Delayed Gamma-Ray Rise}
If, as argued in the previous section, the optical light curve during Phase II (and possibly the end of Phase I) is powered by shocks, it is natural to question why the onset of the $\gamma$-ray emission is modestly delayed with respect to the rise of the optical emission (Figure~1). Particularly puzzling is the extremely rapid rise of the $\gamma$-ray emission, from non-detection to peak luminosity within less than a day. 

One possible explanation for a delay is the finite time is required to accelerate particles up to the required GeV energsies. However, the acceleration time can be estimated as\cite{Metzger+16}
\begin{eqnarray}
&&t_{\rm acc} \approx \frac{Ec}{3e Bv_{\rm f}^{2}} \approx {0.8\,{\rm s}}\,\,\xi\left(\frac{v_{\rm f}}{2000\,{\rm km\,s^{-1}}}\right)^{-5/2}\left(\frac{\epsilon_{B}}{10^{-4}}\right)^{-1/2}\left(\frac{E}{10\,{\rm GeV}}\right)\left(\frac{v_{\rm s}}{400\,{\rm km\,s^{-1}}}\right) \nonumber \\
&& \times \left(\frac{t}{\rm 1\, day}\right) \left(\frac{\dot{M}_{\rm f}}{10^{-4}\,{\rm M_{\odot}\, yr^{-1}}}\right)^{-1/2}, \nonumber \\
\end{eqnarray}
where $B = (6\pi \epsilon_B \mu m_p n_{\rm f}v_{\rm f}^{2})^{1/2}$ is the post-shock magnetic field, where we have taken the density of the upstream gas as that of the fast wind since the reverse shock dominates the total power dissipated by the shocks. This timescale is clearly much too short to explained any observed $\gamma$-ray delay. A similar estimate shows that the pion loss timescale is also too short to explain the observed delay, taking into account the high densities in the radiatively cooled post-shock gas.

A more likely explanation for the delayed $\gamma$-ray rise is early-time absorption by the external slow ejecta, the same column of gas responsible for suppressing the thermal X-ray emission and determining the rise time of the optical emission. As discussed in $\S\ref{sec:optical}$, the optical emission rises on a timescale of $t_{\rm rise} \sim 1$ day when the shocks reach a radius $r \approx v_{\rm sh}t_{\rm rise}$. Substituting this radius into eq.~\ref{eq:tauKN} we find that the $\gamma$-ray optical depth on the timescale of the optical rise is given by (for $\kappa_{\rm opt} = 0.1$ cm$^{2}$ g$^{-1}$)
\be
\tau_{\gamma} \approx 14\xi^{-1}\left(\frac{v_{\rm s}}{400\,{\rm km\,s^{-1}}}\right)^{-1}\left(\frac{E_{\gamma}}{300\,{\rm MeV}}\right)^{-1}\left(\frac{t}{t_{\rm rise}}\right)^{-1}
\ee
For characteristic parameters $\xi \sim 2$, we see that the $0.1-1$ GeV $\gamma$-rays will still be strongly absorbed ($\tau_{\gamma} \gg 1$) on the same timescale as the optical emission is rising, consistent with a delayed onset of the $\gamma$-rays. However, $\tau_{\gamma}$ will decrease at later times as $\tau_{\gamma} \propto t^{-1}$ and with increasing $\gamma$-ray energy $\propto E_{\gamma}^{-1}$, such that the ejecta will become optically thin even at 100 MeV a few days into the fast wind phase and near the start of Phase III. 

The left panel of Supplementary Figure~4 shows a theoretical calculation of the emergent $\gamma$-ray spectrum of the shock after being attenuated by inelastic downscattering through the slow outflow, shown for several values of the Thomson optical depth ahead of the shock, ranging from $\tau_{\rm T} = 0$ (no absorption) to a deeply embedded shock with $\tau_{\rm T} = 500$. These were calculated by means of a Monte Carlo radiative transfer scheme which follows the interaction of the $\gamma$-ray photons with electrons through inelastic Klein-Nishina scattering. Also shown in the right panel of Supplementary Figure~4 is the escape fraction of the total shock $\gamma$-ray luminosity in the 0.1-1 GeV photon energy range (red points), again as a function of the Thomson optical depth ahead of the shock. This is to be compared to the fractional suppression of the {\it optical} luminosity of the shock (black line), which is caused by adiabatic losses due to expansion within the outflow assuming a shock velocity $v_{\rm sh} \approx 600$ km/s and an optical opacity of the slow wind of $\kappa_{\rm opt} = \kappa_{\rm T}$. 

Based on Supplementary Figure~4, as $\tau_{\rm T}$ decreases from 500 to 62.5 during Phase II as the shocks propagate outwards through the slow outflow, this leads the $\gamma$-ray flux to rise by a factor of $\approx 8$ and optical flux to rise by a factor of $\approx 2.3$. This is consistent with the change in the optical and $\gamma$-ray luminosities between the last $\gamma$-ray upper limit during Phase II and the subsequent $\gamma$-ray peak (Figure~1 and Supplementary Table~2). 

If the $\gamma$-ray onset is indeed controlled by such opacity effects, we might expect that the $\gamma$-ray spectrum will evolve from harder to softer with time due to preferential absorption of the softer photons at early times (Supplementary Figure~4). Although the statistics are poor, we see hints of an increase in the photon index with time (Supplementary Figure~5), though future observations will be required for a confirmation. Spectral softening is notably opposite from the trend that would be expected if the spectral evolution was driven by an increase in the maximum particle energy, which according to eq.~\ref{eq:emax} increases linearly with time as the shock propagates to larger radii. \\


\newpage
\noindent
\textbf{Supplementary Table~1:} \textbf{Blackbody temperature of \nova\ estimated from $\bm{(\bv)_0}$ and spectral continuum fitting.} The second/third column are the observed $V$/$B$-band magnitudes (for photometric sets), the fourth column shows the color indices, which have been corrected for the foreground Galactic extinction with $E(B-V)=0.3424$~mag, and the last column shows the inferred temperatures (to nearest 50 K) based on $(\bv)_0$. 

\noindent
\textbf{Supplementary Table~2:} \textbf{Gamma-ray flux as a function of time in \nova.} The daily photon fluxes and detection significances ($\sigma_{\rm det}$) were measured by Fermi LAT with the unbinned likelihood method. 95\% upper limits were computed for bins with $\mathrm{TS}<4$ based on the Bayesian approach in FSSC. The last three columns are the measurements of those time intervals with consecutive daily $\mathrm{TS}<4$. 

\noindent
\textbf{Supplementary Table~3:} \textbf{Gamma-ray spectral properties of ASASSN-16ma, including the time-resolved fitting results before and after the emission dip around MJD 57705.} The abbreviations PL and PLE in the model column stand for simple power-law (i.e., $dN/dE\propto E^{-\Gamma_\gamma}$) and power law with an exponential cutoff (i.e., $dN/dE\propto E^{-s}\exp(-E/E_c)$), respectively. From the last column, the energy cutoff component improves the fit significantly before the dip, but the improvement started to be insignificant after the dip. This may be due to a real spectral change of the nova system or the lower quality of the after-dip data as the $\gamma$-ray flux decreased. 

\renewcommand{\arraystretch}{0.8}

\begin{table}
\centering 
\caption{}
\begin{tabular}{@{}lccccc}
\hline\\
Time & Database$^\mathrm{a}$ & $V$-band & $B$-band & $(\bv)_0$ & Temperature \\
(MJD) & & (mag) & (mag) & (mag) & (K)\\
\hline\\
57687.06 & AAVSO & $11.44\pm0.03$ & $11.76\pm0.07$ & $-0.02\pm0.08$ & 10300\\
57688.45 & AAVSO & $10.29\pm0.01$ & $10.85\pm0.02$ & $0.22\pm0.02$ & 7650 \\
57688.49 & ARAS & $\cdots$ & $\cdots$ & $\cdots$ & 8050 \\
57688.51 & AAVSO & $10.27\pm0.01$ & $10.80\pm0.01$ & $0.19\pm0.02$ & 7900 \\
57689.46 & AAVSO & $10.07\pm0.02$ & $10.63\pm0.02$ & $0.21\pm0.03$ & 7650 \\
57691.50 & ARAS & $\cdots$ & $\cdots$ & $\cdots$ & 8100 \\
57693.52 & ARAS & $\cdots$ & $\cdots$ & $\cdots$ & 7200 \\
57695.46 & AAVSO & $8.18\pm0.01$ & $8.71\pm0.02$ & $0.19\pm0.02$ & 7850 \\
57696.43 & AAVSO & $8.18\pm0.01$ & $8.69\pm0.01$ & $0.16\pm0.01$ & 8100 \\
57697.37 & VSNET & $8.1$ & $8.6$ & $0.17$ & 8100 \\
57698.53 & AAVSO & $7.22\pm0.01$ & $7.83\pm0.01$ & $0.27\pm0.01$ & 7200 \\
57700.99 & AAVSO & $5.7^\mathrm{b}$ & $6.28\pm0.02$ & $0.24$ & 7500 \\
57701.50 & ARAS & $\cdots$ & $\cdots$ & $\cdots$ & 5800 \\
57704.37 & VSNET & $6.1$ & $6.9$ & $0.48$ & 5950 \\
57704.49 & ARAS & $\cdots$ & $\cdots$ & $\cdots$ & 4800 \\
57704.51 & AAVSO & $6.04\pm0.01$ & $6.95\pm0.01$ & $0.57\pm0.01$ & 5550 \\
\hline\\
\multicolumn{6}{l}{$^\mathrm{a}$ The ARAS observations are spectra. }\\
\multicolumn{6}{l}{$^\mathrm{b}$ No statistical error is shown because the magnitude is visually estimated. }\\
\multicolumn{6}{l}{$^\mathrm{c}$ The reported errors are 1$\sigma$ uncertainties. }\\
\end{tabular}
\end{table}
\clearpage

\begin{table}
\centering 
\caption{}
\begin{tabular}{@{}ccc}
\hline\\
Time & Flux$^a$ & Test Statistic$^b$ \\
(MJD) & ($10^{-7}$\pflux) & (TS)\\
\hline\\
57686.5 & $<$1.4 & 0.0\\
57687.5 & $<$2.3 & 0.2\\
57688.5 & $<$1.4 & 0.0\\
57689.5 & $<$1.6 & 1.4\\
57690.5 & $<$1.9 & 0.0\\
57691.5 & $<$1.1 & 0.0\\
57692.5 & $<$1.3 & 0.0\\
57693.5 & $<$1.2 & 0.0\\
57694.5 & $<$1.3 & 0.0\\
57695.5 & $<$1.1 & 0.0\\
57696.5 & $<$1.7 & 0.0\\
57697.5 & $<$1.5 & 0.3\\
57698.5 & $<$2.2 & 0.0\\
57699.5 & $<$1.3 & 0.0\\

57700.5 & $10.4\pm1.3$ & 199.5\\
57701.5 & $7.5\pm2.0$ & 41.6\\
57702.5 & $7.0\pm1.3$ & 106.2\\
57703.5 & $5.5\pm1.1$ & 77.2\\
57704.5 & $5.2\pm1.1$ & 64.8\\
57705.5 & $1.5\pm0.7$ & 13.8\\
57706.5 & $5.1\pm1.1$ & 52.8\\
57707.5 & $4.0\pm1.2$ & 30.7\\
57708.5 & $6.2\pm1.2$ & 67.3\\
57709.5 & $2.5\pm1.7$ & 4.6\\
57710.5 & $<$4.2 & 0.1\\
57711.5 & $<$8.8 & 1.5\\
57712.5 & $<$7.7 & 2.6\\

57713.5 & $2.5\pm1.7$ & 6.2\\
57714.5 & $3.9\pm2.1$ & 6.2\\
57715.5 & $<$2.5 & 0.0\\
57716.5 & $<$4.8 & 1.2\\
57717.5 & $<$3.0 & 0.0\\
57718.5 & $<$2.7 & 0.0\\
57719.5 & $<$1.8 & 0.0\\
57720.5 & $<$2.4 & 0.0\\
\hline\\
57686--57700 & $<0.2$ & 0.0\\
57710--57713 & $1.8\pm1.2$ & 4.2\\
57715--57721 & $<0.7$ & 0.0\\
\hline\\
\multicolumn{3}{l}{$^\mathrm{a}$ The energy range is 100MeV--300GeV. }\\
\multicolumn{3}{l}{$^\mathrm{b}$ The detection significances can be determined by $\sigma^2\approx \sqrt{\mathrm{TS}}$. }\\
\multicolumn{3}{l}{$^\mathrm{c}$ The reported errors are 1$\sigma$ uncertainties and the upper limits are at 95\% confidence level. }\\
\end{tabular}
\end{table}

\begin{table}
\centering 
\caption{}
\begin{tabular}{@{}ccccccc}
\hline\\
Time Range & Model & Photon Index & Cutoff Energy & $F_\mathrm{0.1-300GeV}$ & $\mathrm{TS}^\mathrm{a}$ & Improvement$^\mathrm{b}$ \\
(MJD) & & ($\Gamma_\gamma$ or $s$) & ($E_c$; GeV) & ($10^{-7}$\pflux) & & ($\sigma$) \\
\hline\\
57700--57709 & PL & $2.11\pm0.05$ & $\cdots$ & $5.9\pm0.5$ & 635 & $\cdots$\\
57700--57709 & PLE & $1.86\pm0.11$ & $5.9\pm2.6$ & $5.4\pm0.5$ & 644 & 3.3\\
\hline\\
57700--57704.5 & PL & $2.05\pm0.06$ & $\cdots$ & $7.2\pm0.7$ & 477 & $\cdots$\\
57700--57704.5 & PLE & $1.71\pm0.14$ & $4.2\pm1.9$ & $6.5\pm0.7$ & 490 & 3.5\\
\hline\\
57704.5--57709 & PL & $2.22\pm0.10$ & $\cdots$ & $4.5\pm0.7$ & 179 & $\cdots$\\
57704.5--57709 & PLE & $2.08\pm0.17$ & $12^{+13}_{-12}$ & $4.2\pm0.7$ & 180 & 0.8\\
\hline\\
\multicolumn{7}{l}{$^\mathrm{a}$ The Test Statistic (TS) values of the PL or PLE fits using the Fermi-LAT data in different time ranges. }\\
\multicolumn{7}{l}{$^\mathrm{b}$ The significances of the improvements of PLE over PL, estimated by likelihood ratio test. }\\
\multicolumn{7}{l}{$^\mathrm{c}$ The reported errors are 1$\sigma$ uncertainties. }\\
\end{tabular}
\end{table}
\clearpage

\renewcommand{\figurename}{Supplementary Figure}

\begin{figure}
\includegraphics[width=\textwidth]{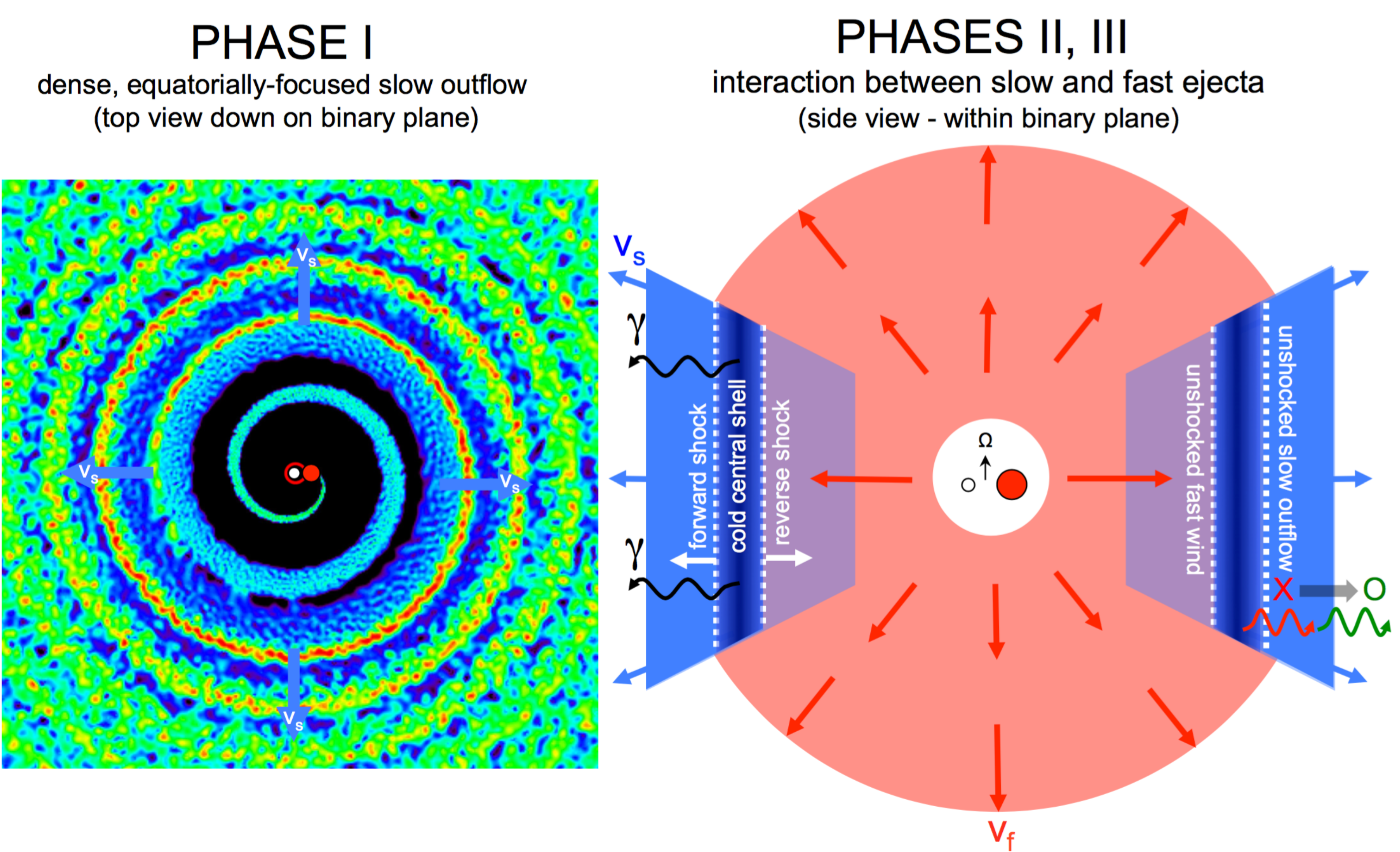}
\centering
\caption{\textbf{Schematic illustration of our proposed model for different phases of ASASSN-16ma.} {\it Left Image:} In Phase I, the nova outburst produces a dense and slow equatorially-focused outflow, possibly due to spiral-like mass loss from the outer L2 Lagrange point of the binary (image from hydrodynamical simulations of Ref.\cite{Pejcha+16a}). {\it Right Image:} In Phase II and III, a faster more spherical outflow (red) expands into the slow outflow produced during Phase I (blue). The collisional interaction between the two flows results in an outwardly propagating forward-reverse shock structure, separated by a cold central shell of gas. Thermal X-rays from the shocks (red lines) are absorbed by the slow outflow ahead of the shocks, reprocessing the shock energy into optical emission (green lines). Relativistic particles accelerated at the shocks are advected into the cold central shell, where they produce non-thermal GeV $\gamma$-ray emission (black lines). Note the change in viewing angle between the left image (top view looking down on the binary plane) and the right image (side view cut through the binary plane).}
\label{fig:cartoon}
\end{figure}

\newpage
\begin{figure}
\includegraphics[width=0.6\textwidth]{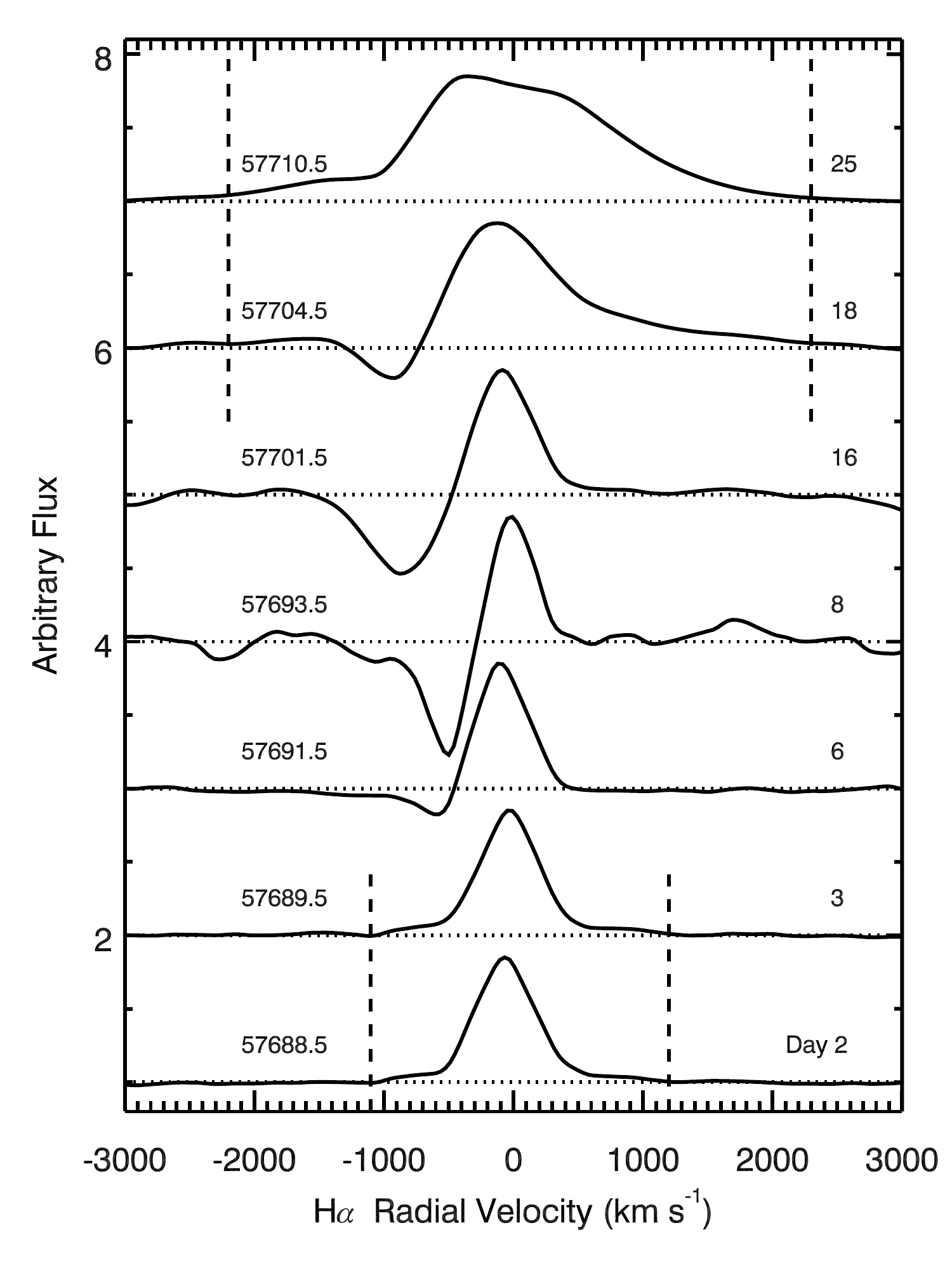}
\centering
\caption{
\textbf{The changing profile of the H$\bm\alpha$ line reveals the evolution of the fast wind and shock.} A time sequence of eight spectra is shown, with the modified Julian date noted to the left and the days since discovery to the right (on this scale, optical maximum is on Day 14). At early times (2--3 days after discovery), H$\alpha$ appears as a narrow, spectrally unresolved component with broad wings extending out to $\pm$1100 km/s. Then a P Cygni absorption trough forms, deepens, and moves blueward. Finally, at 11 days after optical maximum, we see that the H$\alpha$ profile has transformed into a broad emission line with wings extending to $\pm$2200 km/s. We interpret the gradual broadening of the line as the fast component breaking out of an opaque and slower component. A dense, cool shell forms behind the shock between these two phases, and may be the source of the P Cygni absorption. 
}
\end{figure}

\begin{figure}
\includegraphics[width=\textwidth]{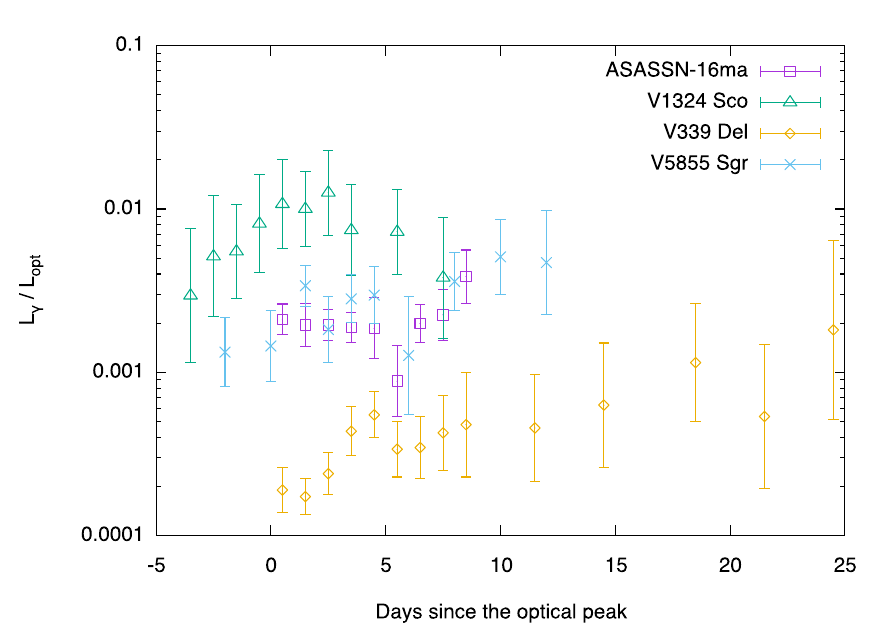}
\centering
\caption{
\textbf{Other Fermi LAT-detected novae show relatively constant ratios of $\bm\gamma$-ray to optical luminosities. }
The ratios of \nova\ and V5855~Sgr were computed using the AAVSO and Fermi-LAT data, while the ratios of V1324~Sco and V339~Del were taken from Ref.\cite{Metzger+15}. The time zeros are set at the optical peaks, which are on June 20, 2012 (V1324~Sco\cite{Cheung+16}), August 16, 2013 (V339~Del\cite{Cheung+16}), October 31, 2016 (V5855~Sgr), and November 8, 2016 (\nova). The reported errors are 1$\sigma$ uncertainties. }
\label{fig:ratio}
\end{figure}

\begin{figure}
\includegraphics[width=\textwidth]{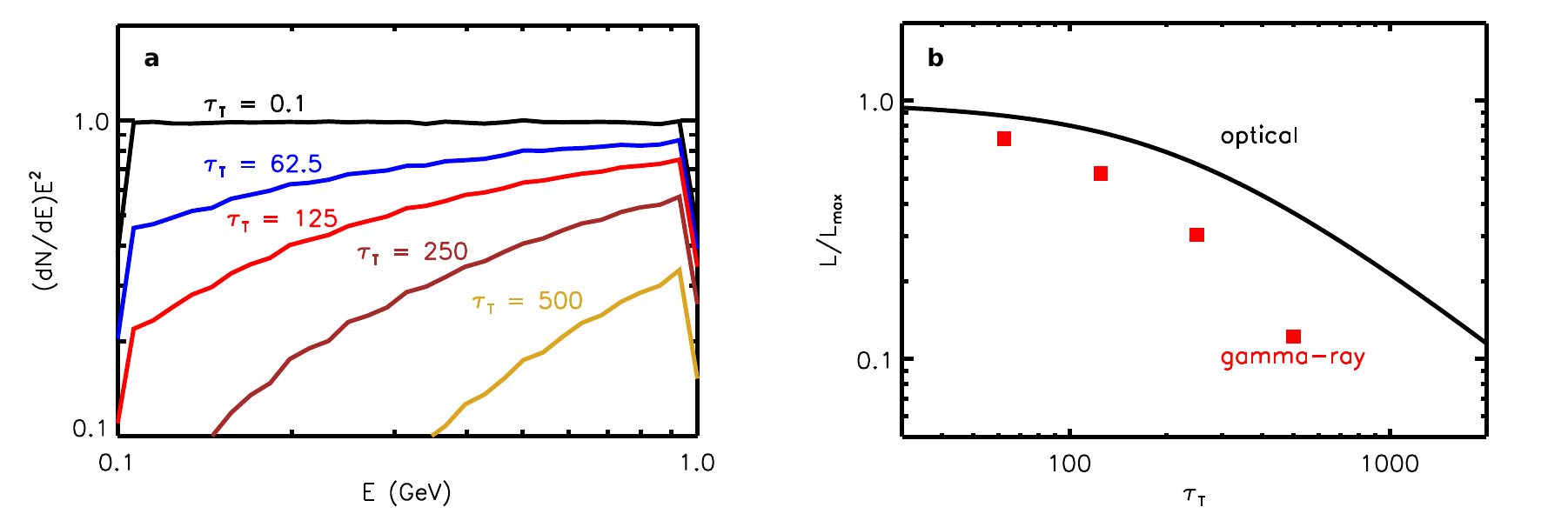}
\centering
\caption{\textbf{As the shocks propagate outwards through the slow outflow, the scattering optical depth $\bm{\tau_{\rm T}}$ ahead of the shock decreases with time, allowing both the optical and $\bm\gamma$-ray emission from the shock to escape unattenuated to the external observer.} {\it Left panel:} $\gamma$-ray spectral energy distribution which emergses from the slow outflow following energy losses due to inelastic down-scattering, when the shock is located at different optical depths $\tau_{\rm T}$ as marked. {\it Right panel:} Red points show the fraction of the total $\gamma$-ray luminosity which escapes at each $\tau_{\rm T}$, i.e. the integral of the attenuated spectrum shown in the left panel across the LAT bandpass. Shown for comparison with a black line is the fraction of the optical luminosity from the shock which escapes when the shock has reached each value of $\tau_{\rm T}$, accounting for PdV adiabatic losses in the expanding medium for an assumed shock expansion velocity of $v_{\rm sh} = 600$ km s$^{-1}$. Both optical and $\gamma$-ray emission rise to their unattenuated value at the same time (i.e. same shock optical depths $\tau_{\rm T} \sim 100$), but the $\gamma$-ray rise is more abrupt, consistent with the rapid onset of the $\gamma$-ray emission in \nova\ around the same time as the optical rise. }
\label{fig:attenuation}
\end{figure}

\begin{figure}
\includegraphics[width=\textwidth]{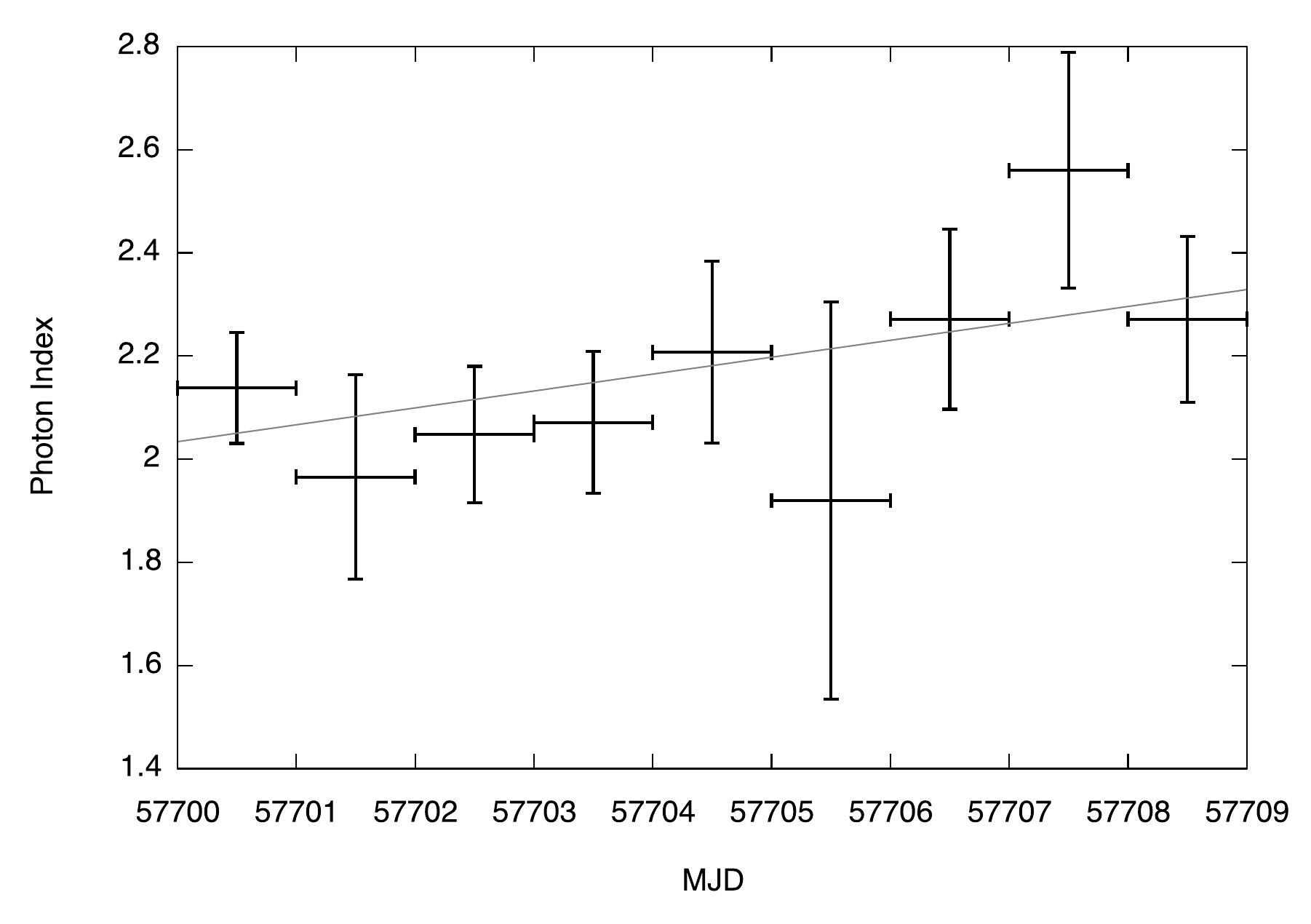}
\centering
\caption{
\textbf{A marginal softening trend is seen in the daily photon index evolution of ASASSN-16ma.} Only the best-fit photon indices (i.e., $\Gamma_\gamma$ in $dN/dE\propto E^{-\Gamma_\gamma}$) with $\mathrm{TS}>10$ are shown. The softening trend can be described by $d \Gamma_\gamma / dt=0.033\pm0.015$ day$^{-1}$ (the solid line) and can be reproduced by the time-resolved analysis presented in Supplementary Table~3. The reported errors are 1$\sigma$ uncertainties. 
}
\end{figure}

\begin{figure}
\includegraphics[width=\textwidth]{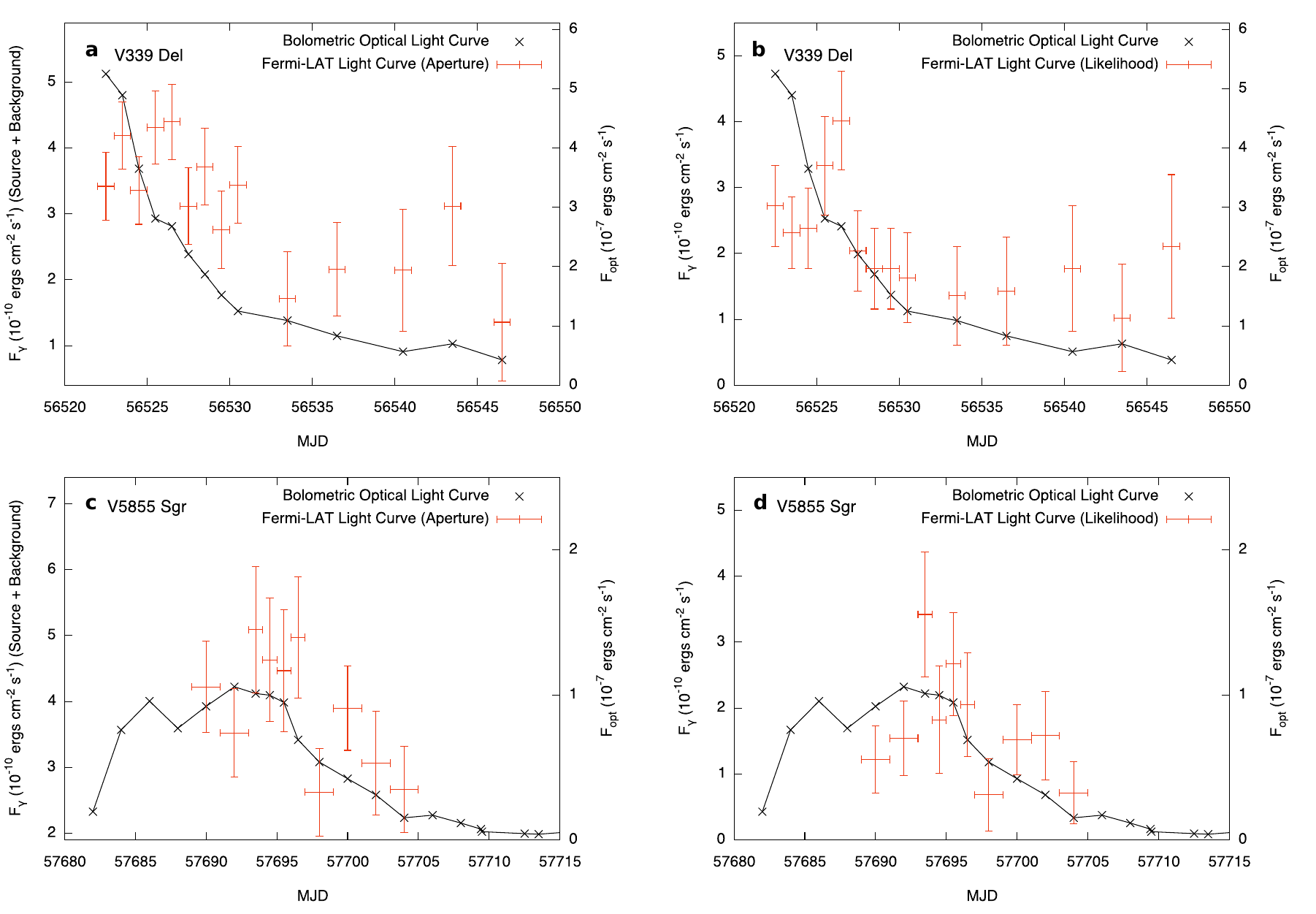}
\centering
\caption{
\textbf{Possible correlations between the $\bm{\gamma}$-ray and optical light curves are also seen in V339 Del and V5855 Sgr, indicating that optical emission from shocks can be common among $\gamma$-ray emitting classical novae besides ASASSN-16ma.} Panels (a) and (b) are the aperture ($2.5\sigma$ correlation) and likelihood ($2.1\sigma$) $\gamma$-ray light curves of V339 Del, respectively, with the bolometric corrected AAVSO optical light curve overplotted. Panels (c) and (d) are the versions for V5855 Sgr, which show a $2.2\sigma$ ($1.8\sigma$) correlation between the aperture (likelihood) $\gamma$-ray and optical light curves. The correlation significances were computed using the same approach as was employed in the case of ASASSN-16ma (see Method). 
For V339 Del, the likelihood $\gamma$-ray light curve and the bolometric corrected optical light curve were obtained from Ref.\cite{Ackermann+14,Metzger+15}. 
For V5855 Sgr, the light curves were extracted in the same way for \nova. However, the binning factor of the likelihood $\gamma$-ray light curve was switched between 1 day and 2 days to optimize the quality of the light curve. In the shown interval of V5855 Sgr, all epochs have $\mathrm{TS}>7$ ($\mathrm{TS_{\rm peak}}=29.1$ on MJD 57695), except for $\mathrm{TS}=1.9$ and 3.0 in MJD 57697--9 and 57703--5, respectively. 
Both the aperture $\gamma$-ray light curves were extracted with a circular region of $r=0.5$ degrees. The reported errors are 1$\sigma$ uncertainties. 
}
\end{figure}

\end{document}